\documentclass[a4paper,fleqn,usenatbib]{mnras}

\usepackage[T1]{fontenc}
\usepackage{ae,aecompl}
\usepackage{textcomp}

\usepackage{graphicx}	
\usepackage{amsmath}	
\usepackage{amssymb}	
\usepackage{multicol}        
\usepackage{bm}		
\usepackage{pdflscape}	
\usepackage{color}          
\usepackage{soul}
\newcommand{\mm}{\mu m}
\newcommand{\gaia}{\it Gaia}
\newcommand{\mz}{$M_{V}-{\rm [Fe/H]}$}
\newcommand{\pkz}{$PM_KZ$}
\newcommand{\pwz}{$PM_{W1}Z$}
\title[RRLs as standard candles in {\it Gaia} DR2]{RR Lyrae stars as standard candles in the {\it \textbf{Gaia}} Data Release 2 Era}
\author[T. Muraveva et al.]{Tatiana Muraveva$^{1}$\thanks{tatiana.muraveva@inaf.it}, Hector E. Delgado$^{2}$, Gisella Clementini$^{1}$, Luis M. Sarro$^{2}$,
\newauthor{Alessia Garofalo$^{1,3}$}
\\
$^{1}$ INAF-Osservatorio di Astrofisica e Scienza dello Spazio di Bologna, Via Piero Gobetti, 93/3, Bologna 40129, Italy\\
$^{2}$ Dpto. de Inteligencia Artificial, UNED, c/ Juan del Rosal, 16, Madrid 28040, Spain\\
$^{3}$ Dipartimento di Fisica e Astronomia, Universit\'a  di Bologna, Via Piero Gobetti 93/2, Bologna 40129,  Italy\\
}

\date{Accepted . Received ; in original form }

\pubyear{2018}

\begin{document}
\label{firstpage}
\pagerange{\pageref{firstpage}--\pageref{lastpage}}
\maketitle

\begin{abstract}

We present results from the analysis of 401 RR Lyrae stars (RRLs)
belonging to the field of the Milky Way (MW). For a  fraction of them multi-band
($V$, $K_\mathrm{s}$, $W1$) photometry, metal abundances, extinction
values and pulsation periods are available in the literature and
accurate trigonometric parallaxes measured by the {\it Gaia} mission alongside {\it Gaia} $G$-band
time-series photometry
have become available with the {\it Gaia} second data release (DR2) on
2018 April 25.
Using a Bayesian fitting approach we derive new near-, mid-infrared
period-absolute magnitude-metallicity ($PMZ$) relations and new
absolute magnitude-metallicity relations in the visual ({\mz}) and $G$ bands ($M_G - {\rm [Fe/H]}$),
 based on the
{\gaia} DR2 parallaxes. We find the dependence of
luminosity on metallicity to be higher than usually found in the
literature, irrespective of the passband considered. Running the
adopted Bayesian model on a simulated dataset we show that the high
metallicity dependence is not caused by the method, but likely arises
from the actual distribution of the data and the presence of a
zero-point offset in the {\it Gaia} parallaxes.
We infer a zero-point offset  of $-0.057$ mas, with the {\it Gaia} DR2
parallaxes being systematically smaller. 
We find the RR Lyrae absolute magnitude  in the $V$, $G$, $K_{\rm s}$ and $W1$ bands at metallicity of [Fe/H]=$-1.5$~dex and period of $P=0.5238$~days, based on {\it Gaia} DR2 parallaxes to be  $M_V=0.66\pm0.06$~mag, $M_G=0.63\pm0.08$~mag, $M_{K_{\rm s}}=-0.37\pm0.11$~mag  and $M_{W1} = -0.41\pm0.11$~mag, respectively. 

\end{abstract}

\begin{keywords}
Parallaxes -- stars: variables: RR Lyrae -- galaxies: Magellanic Clouds -- galaxies: distance
\end{keywords}


\section{Introduction}

Over the years, many different methods have been devised in order to
measure distances in astronomy. However, techniques based on
geometrical principles, among which the trigonometric parallax in
first place, remain the most direct, simple and reliable tool to
anchor the whole astronomical distance ladder on a solid basis.  In
the distance ladder approach, the limited horizon allowed by
parallaxes is circumvented by making use of standard candles, such as,
the RR Lyrae (RRL) variable stars, whose absolute calibration rests on
parallax measurements of local samples of the class.

RR Lyrae stars (RRLs) are old (age $>10$ Gyr), low mass ($< $ 1
M$_{\odot}$), radially pulsating stars which populate the classical
instability strip region of the horizontal branch (HB) in the
colour-magnitude diagram (CMD). RRLs divide into fundamental (RRab)
and first-overtone (RRc) mode pulsators and double-mode (RRd)
variables, which pulsate in both modes simultaneously.  RRLs serve as
standard candles to measure distances since they conform to relations
between the absolute visual magnitude and the metallicity ($M_{V}-{\rm
  [Fe/H]}$), and near- and mid-infrared period-absolute magnitude
($PM$) and $PM$-metallicity ($PMZ$) relations. The near-infrared
$PM_{K}Z$ relation has a number of advantages in comparison with the
visual ${M_{V}-{\rm [Fe/H]}}$ relation, such as a smaller dependence
of the luminosity on interstellar extinction ($A_K$=0.114$A_V$,
\citealt{Cardelli1989}), metallicity and evolutionary effects. 
  The latter cause an intrinsic spread of the {\mz} relation of about
  $\sim 0.2$~mag, while the intrinsic dispersion of the $PMZ$ relation
  due to evolutionary effects is only 0.04 mag in the $K$ band
  \citep{Marconi2015}.

The effect of extinction is even less pronounced in the mid-infrared
passbands. For instance, the extinction in the {\it Wide-field
  Infrared Survey Explorer (WISE)} $W1$ (3.4~$\mu$m) passband is
roughly 15 times smaller than in the $V$ band ($A_{W1}=0.065A_V$,
\citealt{Madore2013}).  Furthermore, near- and mid-infrared light
curves of RRLs have smaller amplitudes, hence, determination of the
mean magnitudes is easier and more precise than in visual bands
further beating down the dispersion of their fundamental
relations. Accurate trigonometric parallaxes for a significantly large
sample of RRLs are needed to firmly calibrate their visual $M_{V}-{\rm
  [Fe/H]}$ and near- and mid-infrared $PMZ$ relations. This is what
{\it Gaia}, a European Space Agency (ESA) cornerstone mission launched
on 2013 December 19, is deemed to provide within a few years
time-frame.

{\it Gaia} is measuring trigonometric parallaxes, positions, proper motions, photometry and main physical parameters for over a billion stars in the Milky Way (MW) and beyond (\citealt{Prusti2016}; \citealt{Brown2016}). The {\it Gaia} first data release (DR1),  on 2016 September 14,  published positions, parallaxes and proper motions for about 2 million stars in common between {\it Gaia} and the Hipparcos and Tycho-2 catalogues, computed as part of the Tycho-{\it Gaia} Astrometric Solution (TGAS, \citealt{Lindegren2016}). The DR1 catalogue comprises parallaxes for 364 MW RRLs,  of which a fraction were used by \citet{Clementini2017} to calibrate the  {\mz}, {\pkz} and {\pwz} relations.
On 2018 April 25, the  {\gaia} second data release (DR2), 
has published  positions  and multi-band photometry for $\sim1.7$ billion sources as well as parallaxes and proper motions calculated solely on {\it Gaia} astrometry for $\sim1.3$ billion sources  \citep{Brown2018}. {\gaia} DR2 also published a catalogue of more than $\sim$500,000 variable stars of different types \citep{Holl2018}, that comprises 140,784 RRLs  \citep{Clementini2018} for which main characteristic parameters (period, pulsation mode, mean magnitudes and amplitudes in the {\gaia} $G$, $G_{\rm BP}$ and $G_{\rm RP}$ passbands, extinction and individual metal abundance [Fe/H]) were also released. This provides an enormous contribution to the common knowledge of the variable star population in and beyond the MW  and also makes it possible to re-calibrate their fundamental relations and extend them to the {\it Gaia} passbands.

 A number of independent studies discussing a possible  zero-point offset affecting the {\gaia} DR2 parallaxes 
 appeared recently  in the literature (e.g. \citealt{Arenou2018}; \citealt{Riess2018}; \citealt{Zinn2018}; \citealt{Stassun2018}). All these studies agree on the {\it Gaia} DR2 parallaxes being  systematically smaller  (hence, providing systematically larger distances) than inferred by other independent techniques. 

 
 In this paper  we use the accurate parallaxes available with {\gaia} DR2  for a large sample of local RRLs  along with a Bayesian fitting approach to derive new {\mz}, {\pkz} and {\pwz} relations,  as well as the $G$-band absolute magnitude-metallicity ($M_G - {\rm [Fe/H]}$) relation.
 The new relations are then used to measure  RRL absolute magnitudes and the distance to RRLs in the Large Magellanic Cloud (LMC), for which the accuracy of the trigonometric parallax measurements is hampered by the faint magnitude/large distance. In doing so we also test the quality of {\gaia} DR2 parallaxes and the zero-point parallax offset.

The paper is organised as follows. In Section~\ref{sec:data} we describe the sample of RRLs  that we have used in this study. In Section~\ref{sec:comp} we perform a comparison of the DR2 parallaxes with the TGAS, Hipparcos and {\it Hubble Space Telescope (HST)} parallaxes and with photometric parallaxes inferred from Baade-Wesselink (BW) studies. The Bayesian fitting approach applied in our study is described in Section~\ref{sec:met}. The RRL {\mz}, {\pkz}, {\pwz} and  $M_G - {\rm [Fe/H]}$
 relations derived in this work and a discussion of the   {\it Gaia} DR2 parallax zero-point offset are presented in Section~\ref{sec:relations}. In Section~\ref{sec:lmc} we use  our newly derived  relations to measure the distance to the LMC.  Finally, a  summary of the paper results and main conclusions are presented in Section~\ref{sec:summ}.

\section{Data}\label{sec:data}
\subsection{The sample of RR Lyrae stars}

In order to calibrate the {\mz}, {\pkz} and {\pwz} relations of RRL variables, one needs a large sample of RRLs with accurate photometry, the precise knowledge of their period and pulsation mode,  metallicities spanning a large enough range, alongside an accurate estimation of the star parallaxes/distances. Following \citet{Clementini2017} we select a sample of 403 MW field RRLs studied by \citet{Dambis2013}, who have collected and homogenised literature values of period, pulsation mode,  extinction in the visual passband ($A_V$), metal abundance ([Fe/H])  and intensity-averaged magnitudes in the Johnson {\it V}, 2MASS $K_\mathrm{s}$ and {\it WISE} $W1$ passbands. \citet{Dambis2013} took the pulsation periods from the ASAS3 catalogue (\citealt{Pojmanski2002}, \citealt{Maintz2005}) and the General Catalogue of Variable Stars  (GCVS, \citealt{gcvs}). The intensity-averaged $V$ magnitudes were calculated from nine overlapping sets of observations (see  \citealt{Dambis2013} and references therein for details); the $K_\mathrm{s}$-band intensity-averaged magnitudes were estimated applying a phase-correction procedure described in  \citet{Feast2008}  to the 2MASS single-epoch $K_\mathrm{s}$ measurements  of \citet{Cutri2003}. \citet{Dambis2013} did not apply phase-corrections to 32 RRLs in their sample. For these objects we  adopted the single-epoch $K_\mathrm{s}$  magnitudes.
 According to figure~B2 in \citet{Feast2008} the largest amplitude of RRLs in the $K_{\rm s}$ band is $\sim0.35$ mag. Hence, for the 32 RRLs with single epoch observations we adopt an uncertainty for the mean $K_{\rm s}$ magnitude of  0.175~mag, corresponding to half the maximum amplitude.
The intensity-averaged $W1$ magnitudes were estimated by \citet{Dambis2013} from  the {\it WISE} single-exposure data. Conforming to the referee request  an additional uncertainty  of 0.02~mag was added to uncertainties in the mean $W1$ magnitudes presented by \citet{Dambis2013}.

\begin{landscape}
\begin{table}
\tiny
\caption{Dataset for the 401 RRLs: (1) name; (2) {\it Gaia} identifier; (3) and (4) coordinates in the {\it Gaia} DR2 catalogue; (5) {\it Gaia} DR2 parallax; (6) RRL type; (7) period; (8), (9) and (10) mean $V$, $K_{\rm s}$ and $W1$ magnitudes from \citet{Dambis2013}; (11) mean $G$ magnitude; (12) metallicity on the \citet{ZW} metallicity scale; (13) extinction in the $V$ band.}{\label{tab:gen}}

\begin{tabular}{l r r r c l l r r r r c l}
\hline
\hline
~~Name   & ${\rm ID}_{\it Gaia}$~~~~~~~~~& RA~~~~ & Dec~~~&  $\varpi_{\rm DR2}$  &Type&~~P & $V~~~~~~~~~$  & $K_{\rm s}$~~~~~~~~ & $W1^{a}$~~~~ & $G$~~~~~~~~~~~& [Fe/H] & ~$A_V$ \\
{} & {} & (deg)~~~~& (deg)~~& (mas) & {} & (day) & (mag)~~~~~~ & (mag)~~~~~~~& (mag)~~~~ & (mag)~~~~~~~~& (dex) & (mag)\\
\hline
   DH    Peg &  2720896455287475584 &  333.85693 &  6.82262   &   2.068 $\pm$ 0.048  &    C$^{b}$ &  0.2555$^{b}$ &   9.553  $\pm$  0.007   & 8.603  $\pm$  0.038    &8.551  $\pm$  0.007   &                      &-1.36  $\pm$  0.20   &  0.263  \\   
   DM    Cyg & 1853751143864356736  & 320.2981   & 32.19129   &   0.965 $\pm$ 0.051  &    AB    &  0.4199$^{b}$ &   11.530 $\pm$  0.018   & 10.287 $\pm$  0.034    &10.228 $\pm$  0.024    & 11.43934 $\pm$  0.00026 &-0.14  $\pm$  0.20   & 0.300   \\   
   DH    Hya & 5737579706158770560  & 135.06169  & -9.779     &   0.469 $\pm$ 0.040  &    AB    &  0.4890$^{b}$ &   12.152 $\pm$  0.009   & 11.143 $\pm$  0.039    &11.042 $\pm$  0.028    & 12.11165 $\pm$  0.00030 &-1.55  $\pm$  0.20   & 0.161   \\   
   DD    Hya & 3090871397797047296  & 123.13255  & 2.83469    &   0.481 $\pm$ 0.051  &    AB    &  0.5018$^{b}$ &   12.202 $\pm$  0.012   & 11.210 $\pm$  0.038    &11.096 $\pm$  0.026    & 12.10446 $\pm$  0.00024 &-1.00  $\pm$  0.20   & 0.000   \\   
   ER    Aps & 5800912537991603200  & 265.99488  & -76.24484  &   0.367 $\pm$ 0.015  &    AB    &  0.4311$^{b}$ &   13.602 $\pm$  0.018   & 11.226 $\pm$  0.037    &11.125 $\pm$  0.006    & 13.32635 $\pm$  0.00009 & -1.39  $\pm$  0.20  &1.100    \\   
   DN    Aqr & 2381771781829913984  & 349.82189  & -24.21641  &   0.642 $\pm$ 0.051  &    AB    &  0.6338$^{b}$ &   11.182 $\pm$  0.008   & 9.901  $\pm$  0.037    &9.897  $\pm$  0.022   &                      &-1.63  $\pm$  0.20   & 0.077   \\   
   BV    Aqr & 6820039248616386688  & 330.72498  & -21.52568  &   0.889 $\pm$ 0.053  &    C     &  0.3638     &   10.888 $\pm$  0.036   & 10.010 $\pm$  0.039    &9.975  $\pm$  0.005   &                      &-1.49  $\pm$  0.20   & 0.103   \\    
   AV    Ser & 4410058473777585024  & 240.9243   & 0.59913    &   0.790 $\pm$ 0.033  &    AB    &  0.4876     &   11.484 $\pm$  0.009   & 10.014 $\pm$  0.037    &10.011 $\pm$  0.015   & 11.38748 $\pm$  0.00035& -1.20  $\pm$  0.20  & 0.514   \\    
   VY    Lib & 6262626680568457600  & 237.82077  & -15.75116  &   0.788 $\pm$ 0.051  &    AB    &  0.5340     &   11.724 $\pm$  0.009   & 10.030 $\pm$  0.036    &10.041 $\pm$  0.017   & 11.53776 $\pm$  0.00025& -1.32  $\pm$  0.20  & 0.588   \\    
   AF    Vel & 5360400630327427072  & 163.26066  & -49.90638  &   0.829 $\pm$ 0.029  &    AB    &  0.5275     &   11.389 $\pm$  0.008   & 10.042 $\pm$  0.038    &9.967  $\pm$  0.009   & 11.25960 $\pm$  0.00015& -1.64  $\pm$  0.20  & 0.407   \\    
   BB    Eri & 2976126948438805760  & 73.40647   & -19.43363  &   0.601 $\pm$ 0.030  &    AB    &  0.5701     &   11.498 $\pm$  0.009   & 10.047 $\pm$  0.036    &10.168 $\pm$  0.008   & 11.39119 $\pm$  0.00016& -1.51  $\pm$  0.20  & 0.148   \\    
   SS    Psc & 289662043370304384   & 20.21817   & 21.72867   &   0.821 $\pm$ 0.122  &    C     &  0.2879     &   10.979 $\pm$  0.012   & 10.048 $\pm$  0.036    &9.985  $\pm$  0.005   &                      &-0.82  $\pm$  0.20   & 0.149   \\    
   SW    Aqr & 2689556491246048896  & 318.8242   & 0.07611    &   0.888 $\pm$ 0.048  &    AB    &  0.4594     &   11.176 $\pm$  0.005   & 10.056 $\pm$  0.037    &10.071 $\pm$  0.045   &                      &-1.24  $\pm$  0.20   & 0.233   \\    
   RW    TrA & 5815008831122635520  & 255.19446  & -66.66392  &   1.038 $\pm$ 0.037  &    AB    &  0.3741     &   11.347 $\pm$  0.009   & 10.058 $\pm$  0.036    &10.023 $\pm$  0.006   & 11.22482 $\pm$  0.00015& 0.07   $\pm$  0.20  & 0.416   \\    
   SX    UMa & 1565435491136901888  & 201.55555  & 56.25696   &   0.755 $\pm$ 0.043  &    C     &  0.3072     &   10.859 $\pm$  0.018   & 10.066 $\pm$  0.035    &10.046 $\pm$  0.004   & 10.77856 $\pm$  0.00021& -1.78  $\pm$  0.20  & 0.030   \\    
   BK    Dra & 2254942462734092288  & 289.58597  & 66.41345   &   0.711 $\pm$ 0.025  &    AB    &  0.5921     &   11.169 $\pm$  0.018   & 10.069 $\pm$  0.034    &9.990  $\pm$  0.004   & 11.09659 $\pm$  0.00014& -2.12  $\pm$  0.20  & 0.097   \\    
   XX    Pup & 5721192383002003200  & 122.11754  & -16.53325  &   0.667 $\pm$ 0.034  &    AB    &  0.5172     &   11.237 $\pm$  0.009   & 10.084 $\pm$  0.038    &10.007 $\pm$  0.019   & 11.13463 $\pm$  0.00017& -1.42  $\pm$  0.20  & 0.192   \\    
   BC    Dra & 2269585754295172608  & 273.57913  & 76.68579   &   0.629 $\pm$ 0.019  &    AB    &  0.7196     &   11.588 $\pm$  0.036   & 10.088 $\pm$  0.037    &10.069 $\pm$  0.003   & 11.40802 $\pm$  0.00045& -2.00  $\pm$  0.20  & 0.208   \\    
   VW    Scl & 4985455994038393088  & 19.56251   & -39.21262  &   0.850 $\pm$ 0.073  &    AB    &  0.5110     &   11.029 $\pm$  0.013   & 10.109 $\pm$  0.037    &10.007 $\pm$  0.008   & 11.01849 $\pm$  0.00044& -1.06  $\pm$  0.20  & 0.048   \\    
   Z     Mic & 6787617919184986496  & 319.09467  & -30.28421  &   0.815 $\pm$ 0.066  &    AB    &  0.5870     &   11.612 $\pm$  0.009   & 10.113 $\pm$  0.037    &10.046 $\pm$  0.037   & 11.47175 $\pm$  0.00065& -1.28  $\pm$  0.20  & 0.286   \\    
   X     Crt & 3587566361077304704  & 177.23426  & -10.44142  &   0.580 $\pm$ 0.057  &    AB    &  0.7328     &   11.465 $\pm$  0.006   & 10.148 $\pm$  0.038    &10.106 $\pm$  0.008   & 11.33385 $\pm$  0.00021& -1.75  $\pm$  0.20  & 0.083   \\    
   V0690 Sco & 4035521829393903744  & 269.41099  & -40.5576   &   0.880 $\pm$ 0.040  &    AB    &  0.4923     &   11.419 $\pm$  0.012   & 10.168 $\pm$  0.037    &10.022 $\pm$  0.009   &                      &-1.11  $\pm$  0.20   & 0.375   \\    
   TW    Boo & 1489614955993536000  & 221.27477  & 41.0287    &   0.724 $\pm$ 0.023  &    AB    &  0.5323     &   11.264 $\pm$  0.018   & 10.176 $\pm$  0.036    &10.108 $\pm$  0.004   & 11.17810 $\pm$  0.00019& -1.41  $\pm$  0.20  & 0.041   \\    
   ST    Com & 3940418398550912512  & 199.46383  & 20.78062   &   0.672 $\pm$ 0.034  &    AB    &  0.5990     &   11.438 $\pm$  0.009   & 10.191 $\pm$  0.036    &10.139 $\pm$  0.009   & 11.31836 $\pm$  0.00018& -1.26  $\pm$  0.20  & 0.072   \\    
   EL    Aps & 5801111519533424384  & 263.92191  & -76.22108  &   0.709 $\pm$ 0.024  &    AB    &  0.5798     &   11.896 $\pm$  0.018   & 10.193 $\pm$  0.037    &10.130 $\pm$  0.006   & 11.73753 $\pm$  0.00018& -1.56  $\pm$  0.20  & 0.659   \\    
   AM    Vir & 3604450388616968576  & 200.88886  & -16.66627  &   0.720 $\pm$ 0.041  &    AB    &  0.6151     &   11.525 $\pm$  0.012   & 10.197 $\pm$  0.037    &10.085 $\pm$  0.007   &                      &-1.37  $\pm$  0.20   & 0.205   \\    
   TV    Boo & 1492230556717187456  & 214.1524   & 42.35977   &   0.747 $\pm$ 0.028  &    C     &  0.3126     &   10.999 $\pm$  0.018   & 10.210 $\pm$  0.034    &10.154 $\pm$  0.004   & 10.91106 $\pm$  0.00012& -2.24  $\pm$  0.20  & 0.030    \\   
   AT    Ser & 4454183799545435008  & 238.9182   & 7.98874    &   0.575 $\pm$ 0.046  &    AB    &  0.7465     &   11.492 $\pm$  0.007   & 10.214 $\pm$  0.036    &10.147 $\pm$  0.009   & 11.37944 $\pm$  0.00032& -2.03  $\pm$  0.20  & 0.114    \\   
   CG    Lib & 6238435088295762048  & 233.81972  & -24.33689  &   0.862 $\pm$ 0.046  &    C     &  0.3068     &   11.511 $\pm$  0.012   & 10.224 $\pm$  0.038    &10.127 $\pm$  0.007   &                      &-1.32  $\pm$  0.20   & 0.590$^{b}$\\      
   AP    Ser & 1167409941124817664  & 228.50366  & 9.98089    &   0.749 $\pm$ 0.038  &    C     &  0.3408     &   11.078 $\pm$  0.009   & 10.233 $\pm$  0.038    &10.159 $\pm$  0.005   &                      &-1.61  $\pm$  0.20   & 0.127    \\   
   TW    Her & 4596935593202765184  & 268.63002  & 30.41046   &   0.860 $\pm$ 0.024  &    AB    &  0.3996     &   11.274 $\pm$  0.018   & 10.238 $\pm$  0.034    &10.218 $\pm$  0.006   & 11.20763 $\pm$  0.00029& -0.67  $\pm$  0.20  & 0.172    \\   
   SZ    Hya & 5743059538967112576  & 138.45336  & -9.31929   &   0.775 $\pm$ 0.044  &    AB    &  0.5374     &   11.277 $\pm$  0.018   & 10.255 $\pm$  0.038    &10.147 $\pm$  0.012   &                      &-1.75  $\pm$  0.20   & 0.114    \\   
   V1645 Sgr & 6680420204104678272  & 305.18548  & -41.11846  &   0.622 $\pm$ 0.048  &    AB    &  0.5530     &   11.378 $\pm$  0.036   & 10.258 $\pm$  0.038    &10.170 $\pm$  0.007   &                      &-1.74  $\pm$  0.20   & 0.173    \\   
   ST    Oph & 4370549580720839296  & 263.49738  & -1.08085   &   0.762 $\pm$ 0.049  &    AB    &  0.4504     &   12.184 $\pm$  0.007   & 10.261 $\pm$  0.038    &10.474 $\pm$  0.011   & 11.85780 $\pm$  0.00146& -1.30  $\pm$  0.20  & 0.832    \\   
   RX    Cet & 2373827054405340800  & 8.40938    & -15.4877   &   0.648 $\pm$ 0.078  &    AB    &  0.5735     &   11.428 $\pm$  0.009   & 10.277 $\pm$  0.037    &10.163 $\pm$  0.016   & 11.27232 $\pm$  0.00030& -1.46  $\pm$  0.20  & 0.075    \\   
   RR    Gem & 886793515494085248   & 110.38971  & 30.88319   &   0.687 $\pm$ 0.047  &    AB    &  0.3973     &   11.369 $\pm$  0.012   & 10.279 $\pm$  0.036    &10.211 $\pm$  0.007   &                      &-0.35  $\pm$  0.20   & 0.238    \\   
   AA    CMi & 3111925220109675136  & 109.32986  & 1.72779    &   0.756 $\pm$ 0.045  &    AB    &  0.4764     &   11.552 $\pm$  0.018   & 10.287 $\pm$  0.036    &10.221 $\pm$  0.010   & 11.49118 $\pm$  0.00020& -0.55  $\pm$  0.20  & 0.257    \\   
   ST    Leo & 3915998558830693888  & 174.63612  & 10.56144   &   0.749 $\pm$ 0.061  &    AB    &  0.4780     &   11.516 $\pm$  0.007   & 10.290 $\pm$  0.038    &10.389 $\pm$  0.018   &                      &-1.29  $\pm$  0.20   & 0.115    \\   
   V0674 Cen & 6120897123486850944  & 210.85004  & -36.4057   &   0.812 $\pm$ 0.072  &    AB    &  0.4940     &   11.276 $\pm$  0.018   & 10.297 $\pm$  0.040    &10.102 $\pm$  0.005   &                      &-1.53  $\pm$  0.20   & 0.198    \\   
   V0494 Sco & 4055098870077726976  & 265.20198  & -31.54219  &   0.923 $\pm$ 0.052  &    AB    &  0.4273     &   11.330 $\pm$  0.009   & 10.330 $\pm$  0.042    &                    &                      &-1.01  $\pm$  0.20   & 0.588    \\   
   AT    Vir & 3677686044939929728  & 193.79366  & -5.45907   &   0.840 $\pm$ 0.077  &    AB    &  0.5258     &   11.335 $\pm$  0.012   & 10.337 $\pm$  0.042    &10.203 $\pm$  0.010   & 11.41762 $\pm$  0.00056& -1.91  $\pm$  0.20  & 0.092    \\   
   W     Tuc & 4709830423483623808  & 14.5405    & -63.39574  &   0.566 $\pm$ 0.026  &    AB    &  0.6423     &   11.433 $\pm$  0.008   & 10.340 $\pm$  0.037    &10.304 $\pm$  0.015   & 11.35078 $\pm$  0.00018& -1.64  $\pm$  0.20  & 0.063    \\   
   AO    Tuc & 4918030715504071296  & 1.02651    & -59.48524  &   0.732 $\pm$ 0.028  &    C     &  0.3333     &   11.107 $\pm$  0.018   & 10.364 $\pm$  0.175$^{c}$&10.298 $\pm$  0.006   & 11.05707 $\pm$  0.00011& -1.25  $\pm$  0.20  & 0.031    \\   

\hline 
\end{tabular}

 $^{a}$ An additional uncertainty of 0.02~mag was added to the uncertainties in the mean $W1$ magnitudes presented in the table while running the model. See text for the details.\\
$^{b}$ Parameters were changed according to updated informations. See text for the details.\\
$^{c}$ For the 32 RRLs with singe epoch measurements in the $K_{\rm s}$ band an uncertainty for the mean $K_{\rm s}$ magnitudes of 0.175~mag was assumed. See text for the details.\\ 
This table is published in its entirety at the CDS;  a portion is shown here for guidance regarding its form and content. 
\end{table}
\end{landscape}

 The uncertainties in periods of RRLs were considered to be 1\% in their decadic logarithm. Extinction values were inferred from the three-dimensional model of \citet{Drimmel2003}  derived from the dust emission maps of \citet{SFD1998}. Individual uncertainties of the extinction values are not provided by \citet{Dambis2013}, hence, we adopt the reddening uncertainties of  $0.16*E(B-V)$ as suggested by \citet{SFD1998} for RRLs in our sample.
The $V$, $K_{\rm s}$ and $W1$ apparent magnitudes were corrected for interstellar extinction adopting $R_V$ = 3.1, $A_K /A_V = 0.114$ (\citealt{Cardelli1989}, \citealt{Caputo2000a}) and $A_{W1} /A_{V }= 0.065$ \citep{Madore2013}.  \citet{Dambis2013} calculated homogeneous metallicities on the \citet{ZW} metallicity scale, combining spectroscopically and photometrically measured metal abundances.  Uncertainties of  individual metallicities are not  provided in the  \citet{Dambis2013} catalogue. We assumed them to be of 0.1 dex for the stars that have metallicity estimates from  high-resolution spectroscopy. An uncertainty  of 0.2~dex was instead adopted for RRLs whose metal abundance was  measured with the $\Delta S$ technique \citep{Preston1959} or for which we have not found the source of the metallicity estimate. Finally, we assigned a metallicity uncertainty of 0.3~dex to all stars, whose metallicity was obtained from photometry or other non-spectroscopic methods.  

While working on this paper we became aware of updated parameter values for some RRLs in our sample (J.~Lub, private communication). Following these updates,  we adopted a different period value than in \citet{Dambis2013} for six RRLs  (namely, DH~Peg, ER~Aps, DN~Aqr, DM~Cyg, DD~Hya, DH~Hya) and the pulsation mode of DH Peg was changed to RRc, accordingly.
The extinction values of CG~Lib and RZ~Cep were also revised. Finally, BB~Vir turned out to be a blend of two stars and was hence discarded. On the other hand, the rather long period of BI~Tel (P=1.17 days), would place the star in the Anomalous Cepheid domain. Hence, we decided to discard this RRL as well. Our final sample consists of 401 RRLs, 
 of which 366 pulsate in the fundamental mode and 35 in the first-overtone mode. 
 
 We have crossmatched our catalogue of 401 RRLs against the {\gaia} DR2 catalogue available through the {\it Gaia}  Archive website\footnote{\url{http://archives.esac.esa.int/gaia}} using a cross-match radius of  4\arcsec,  and recovered the DR2 parallaxes  for all of them. The complete dataset, namely, identifications, parallaxes, positions and mean $G$ magnitudes from  the {\gaia} DR2 catalogue, alongside  the period, pulsation mode, extinction, metal abundance and mean $V$, $K_\mathrm{s}$ and $W1$ magnitudes available in the literature  for these 401 RRLs, are provided in Table~\ref{tab:gen}.  The parallaxes of our sample span the range from $-2.61$~mas to 2.68~mas, with seven RRLs having a negative parallax value, among which, unfortunately, is RR Lyr itself, the bright RRL that gives its name to the whole class (see Section~\ref{sec:comp-paral}). Uncertainties of the {\gaia} DR2 parallaxes for the 401 RRLs in our sample are shown by the red histogram in Figure~\ref{fig:hist_err}. They range from 0.01 to 0.61~mas. 
The position on sky of the 401 RRLs 
is shown in Fig.~\ref{fig:map}. They appear to be homogeneously distributed all over the sky, which makes any possible systematic spatially-correlated biases negligible. The apparent $V$ mean magnitudes of the 401 RRLs range  
from 7.75 to 16.81~mag. Adopting for the RRL mean $V$absolute magnitude $M_V$=0.59~mag at [Fe/H]=$-1.5$~dex \citep{Cacciari2003} we find for our sample distance moduli  spanning the  range from $\sim$ 7 to $\sim$16~mag or distances from $\sim250$ to $\sim16,000$ pc. Periods and metallicities of the 401 RRLs also span quite large ranges, namely, from 0.25 to 0.96~days in period and from $-2.84$ to +0.07~dex in metallicity, with the metallicity distribution of the sample peaking at [Fe/H]$\sim - 1.5$~dex. The distributions in apparent $V$ mean magnitude, period and metallicity of our sample of 401 RRLs are shown in the upper, middle and lower panels of Fig.~\ref{fig:hist_all}, respectively. 

Regarding a possible selection bias, our sample  is mainly affected by the
selection process carried out in \citet{Dambis2013}. The requirements set
there have effects potentially stronger than most of the {\it Gaia}
selection function characteristics \citep[described qualitatively
in][and references therein]{Brown2018}. In
particular, \citet{Dambis2013} require that the stars in their sample have
metallicity and distance estimates. This in general results in a
global overrepresentation of intrinsically brighter stars. This
overrepresentation may be negligible for nearby stars, but will become
significant for the most distant stars which are predominantly
metal-poor halo stars. On the contrary, we expect no selection effect
in period except those that may arise as a consequence of indirect
correlations with absolute magnitude.
 
This large sample of homogeneously distributed MW RRLs whose main characteristics span significantly large ranges in parameter space, in combination with the DR2 accurate parallaxes (see Fig.~\ref{fig:hist_err})  and $G$-band photometry allows us to study with unprecedented details  the  infrared $PM$ and $PMZ$, and the visual {\mz} relations of RRLs and to derive for the first time the $M_{G} - {\rm [Fe/H]}$ relation in the {\it Gaia} $G$ band (see Section~\ref{sec:relations}).

\section{Comparison with literature data}\label{sec:comp}
\subsection{Comparison with previous parallax estimates in the literature}\label{sec:comp-paral}
The lack of accurate trigonometric parallaxes for a significantly large sample of RRLs has been so far a main limitation hampering the use of RRLs as standard candles of the cosmic distance ladder. 
The ESA mission {\it Hipparcos} (\citealt{vanLeeuwen2007}  and references therein)  measured the trigonometric parallax 
of more than a hundred RRLs, however,   for the vast majority of them the parallax uncertainty is larger than $\sim 30$\%. Trigonometric parallaxes measured  with the Fine Guide Sensor (FGS) on board  the {\it HST} have been published by \citet{Benedict2011} for only five MW RRLs: RZ~Cep, SU~Dra, UV~Oct, XZ~Cyg and RR~Lyr itself. Finally, with {\it Gaia} DR1 in 2016 trigonometric parallaxes calculated as part of the TGAS were made available for 364 RRLs. 
In this section we compare the recently published  {\it Gaia} DR2 parallaxes with the RRL parallax measurements available so far.

\begin{figure}
   \includegraphics[trim=0 20 0 0,width=\linewidth]{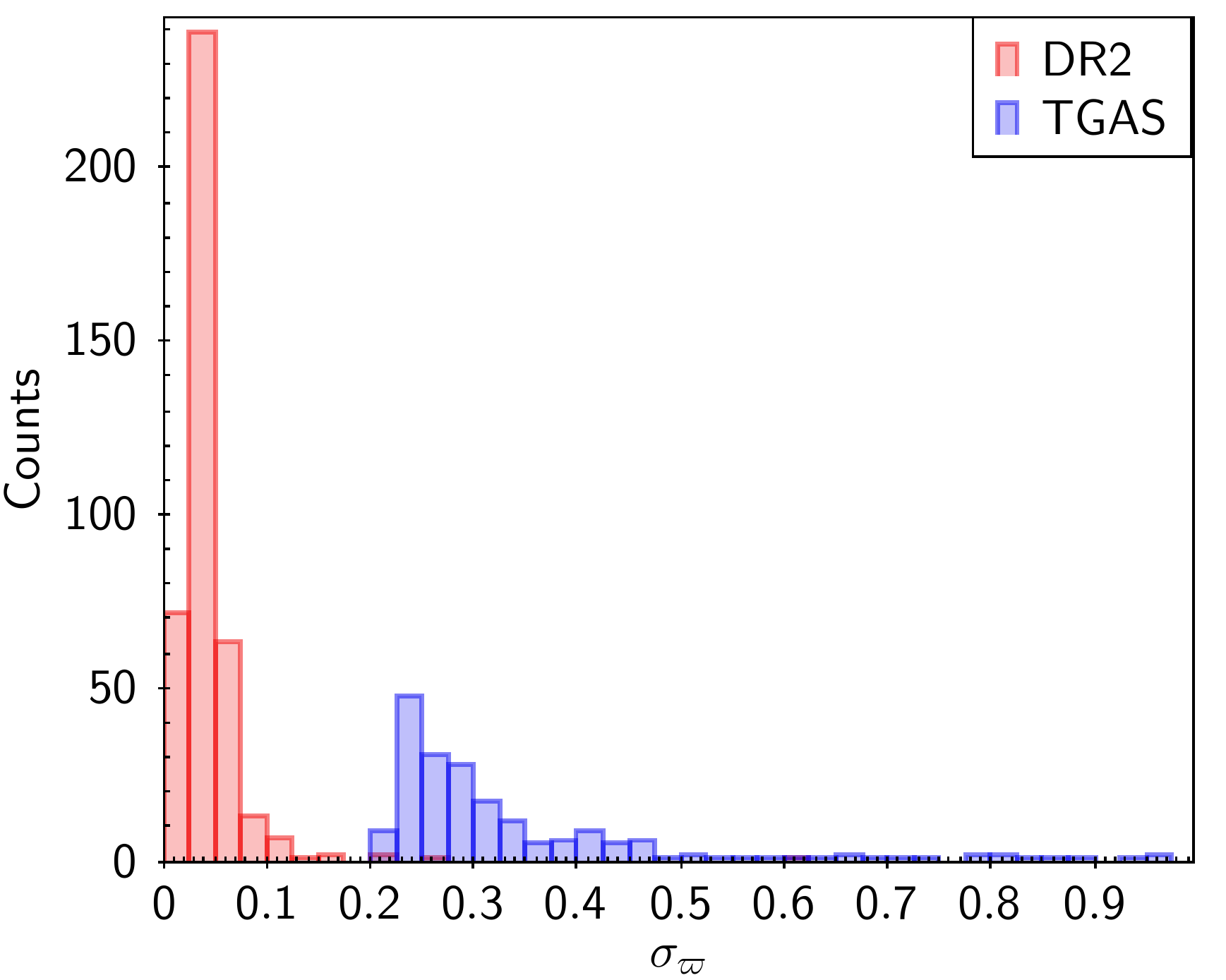}
  \caption{Distribution of the DR2 parallax uncertainties for our sample of 401 RRLs (red histogram) and the TGAS parallax uncertainties for the subsample of 199 RRLs for which TGAS parallaxes were published in {\it Gaia} DR1 (blue histogram). The bin size is 0.025~mas.}
  \label{fig:hist_err}
\end{figure}

\begin{figure}
     \includegraphics[trim=0 20 0 20,width=\linewidth]{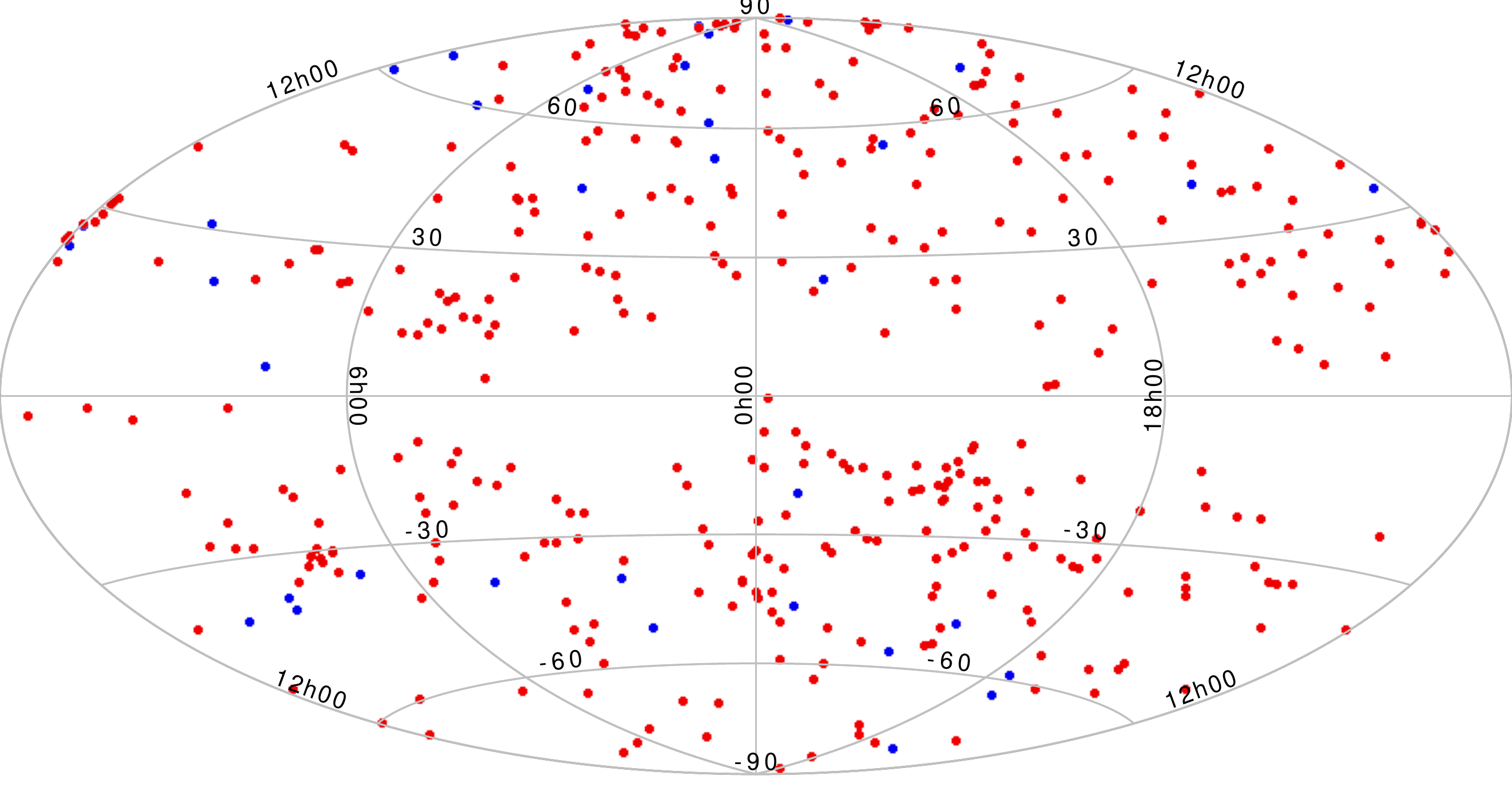}
  \caption{Sky distribution of our sample of 401 RRLs  in Galactic coordinates. Red and blue filled circles show RRab and RRc stars respectively.}
  \label{fig:map}
\end{figure}

TGAS parallaxes are available in {\it Gaia} DR1 for 199 of the RRLs in our sample of 401. The blue histogram in Fig.~\ref{fig:hist_err}  shows the distribution of uncertainties of the TGAS parallaxes for these 199 RRLs. The reduced uncertainty of the DR2 parallaxes (red histogram) with respect to TGAS is impressive.   The difference between DR2 and TGAS parallaxes plotted versus DR2 parallax values is shown in Fig.~\ref{fig:comp_tgas_dr2}. The DR2 parallaxes are generally in reasonably good agreement with the TGAS estimates, except for RR Lyr itself.  The DR2 parallax of RR Lyr has a large negative value ($-2.61 \pm 0.61$~mas) and deviates significantly from  the TGAS parallax estimate ($3.64 \pm 0.23$~mas), hence,  we do not plot the star in Fig.~\ref{fig:comp_tgas_dr2}. The wrong DR2 parallax for RR Lyr was caused by an incorrect estimation of the star's mean $G$ magnitude (17.04~mag, which is $\sim$ 10 mag fainter than the star true magnitude), that induced an incorrect estimation of the magnitude-dependent term applied in the astrometric instrument calibration (\citealt{Arenou2018}, \citealt{Brown2018}). 

\begin{figure}
      \includegraphics[trim=0 40 0 0,width=1.05\linewidth]{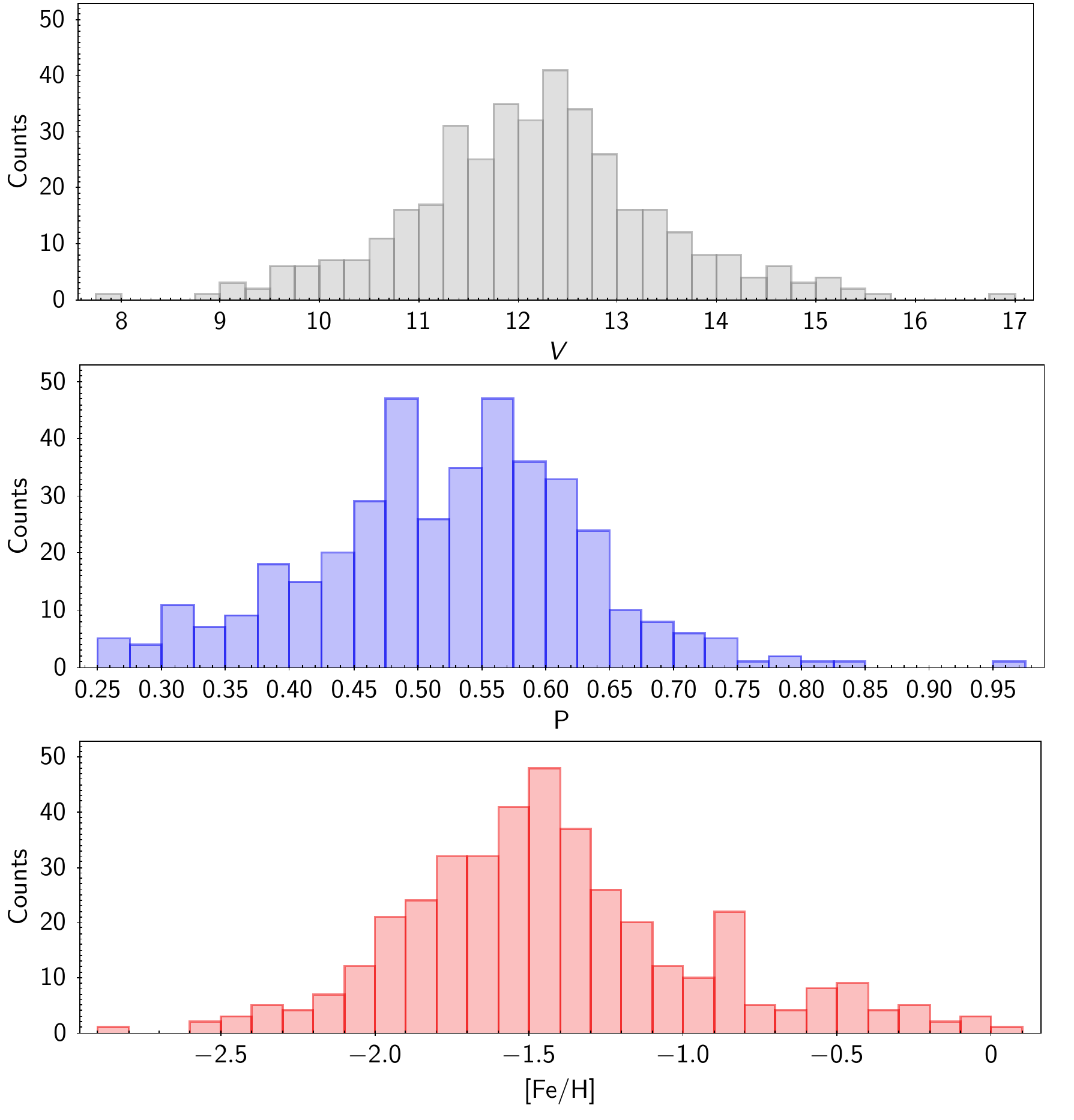}
  \caption{Distribution in apparent $V$ mean magnitude ({\it upper panel}), period ({\it middle panel}) and metallicity (on the \citealt{ZW} metallicity scale; {\it lower panel}) for the RRLs in our sample.  $V$ mean magnitudes are available for 382  RRLs in our sample of 401 variables, metallicities for 400 RRLs, while pulsation periods are available for the whole sample of 401 stars.}
  \label{fig:hist_all}
\end{figure}

A weighted  least-squares fit of the relation $\varpi_{TGAS}=\alpha\varpi_{DR2}$ returns a slope $\alpha=1.02\pm0.02$, which is close to the bisector line slope. However, there is a significant spread  for $\varpi_{DR2}\sim0.3 - 1$~mas in Fig.~\ref{fig:comp_tgas_dr2}, with the TGAS parallaxes having negative or significantly larger values than the DR2 parallaxes.
 The  non-weighted mean difference between DR2 and TGAS parallaxes: $\Delta \varpi_0 = \varpi_{\rm DR2} - \varpi_{\rm TGAS}$, omitting RR Lyr itself, is $-0.04$~mas.
However, the large uncertainties of the TGAS parallaxes prevent a reliable estimation of any zero-point offset that might exist between the DR2 
and the TGAS parallaxes of RRLs.

Table~\ref{tab:varhst}  shows the comparison  for five RRLs for which Hipparcos, {\it HST}, TGAS and {\it Gaia} DR2 parallax measurements are available.
There is a general agreement between the {\it HST}, TGAS and DR2 parallaxes except for RR Lyr. 
Fig.~\ref{fig:comp_parallax_all} shows the Hipparcos (lower panel), {\it HST} (middle panel) and TGAS (upper panel) parallaxes plotted versus {\gaia} DR2 parallaxes for 4 of those five RRLs. For the sake of clarity we did not plot  RR Lyr in the figure. Similarly to Fig.~\ref{fig:comp_tgas_dr2} the upper panel of Fig.~\ref{fig:comp_parallax_all} shows the  nice agreement existing between TGAS and DR2 parallaxes. Agreement between {\gaia} DR2 and Hipparcos parallaxes (lower panel) is less pronounced; on the contrary, a very nice agreement of the DR2 and {\it HST} parallaxes is  seen in the middle panel of Fig.~\ref{fig:comp_parallax_all}, confirming the  reliability of the {\it Gaia} DR2 parallaxes. However, the sample of RRLs with both, DR2 and {\it HST} parallax estimates available is too small to measure any possible zero-point offset of the {\gaia} parallaxes  with respect to {\it HST}. The interested reader is referred to \citet{Arenou2018} who validated {\gaia} DR2 catalogue and found a negligible ($-0.01\pm0.02$~mas) offset between the {\it HST} and DR2 parallaxes using a sample of stars significantly larger than the few RRLs that could be used here.



\begin{table*}
\caption[]{Hipparcos, {\it HST}, TGAS and {\it Gaia} DR2 parallaxes of RRLs for which these independent measurements are available.\label{tab:varhst}}
\begin{center}
\begin{tabular}{l c l c r}
\hline
\hline
Name   & $\varpi_{Hipparcos}$ & ~~~$\varpi_{HST}$ &$\varpi_{\rm TGAS}$ &  $\varpi_{\rm DR2}$~~~~\\
   &(mas) & ~~~~(mas) &  (mas) & (mas)~~~~\\
\hline
RR Lyr & $3.46 \pm 0.64$  & $3.77 \pm 0.13$    &    $3.64 \pm 0.23$  & $-2.61 \pm 0.61$ \\ 
RZ Cep  & $0.59 \pm 1.48$ & $2.12  \pm 0.16*$  & $2.65 \pm 0.24$    & $2.36 \pm 0.03$  \\
SU Dra &  $0.20 \pm 1.13$ &  $1.42 \pm 0.16$ & $1.43 \pm 0.28$   &  $1.40 \pm 0.03$ \\
UV Oct &  $2.44 \pm 0.81$   & $1.71 \pm 0.10$  &   $2.02 \pm 0.22$ & $1.89 \pm 0.03$    \\
XZ Cyg  &   $2.29 \pm 0.84$  &   $1.67 \pm 0.17$ & $1.56 \pm 0.23$ &  $1.57 \pm 0.03$      \\
\noalign{\smallskip}
\hline 
\end{tabular}
\end{center}
* \citet{Benedict2011} provide  two values for the parallax of RZ Cep: $\varpi = 2.12$~mas and $\varpi = 2.54$~mas, with  the former being the final and preferred adopted value (according to Benedict  private communication in \citealt{Monson2017}). 
\\
\normalsize
\end{table*}


\begin{figure}
     \includegraphics[trim=30 200 30 130,width=1.05\linewidth]{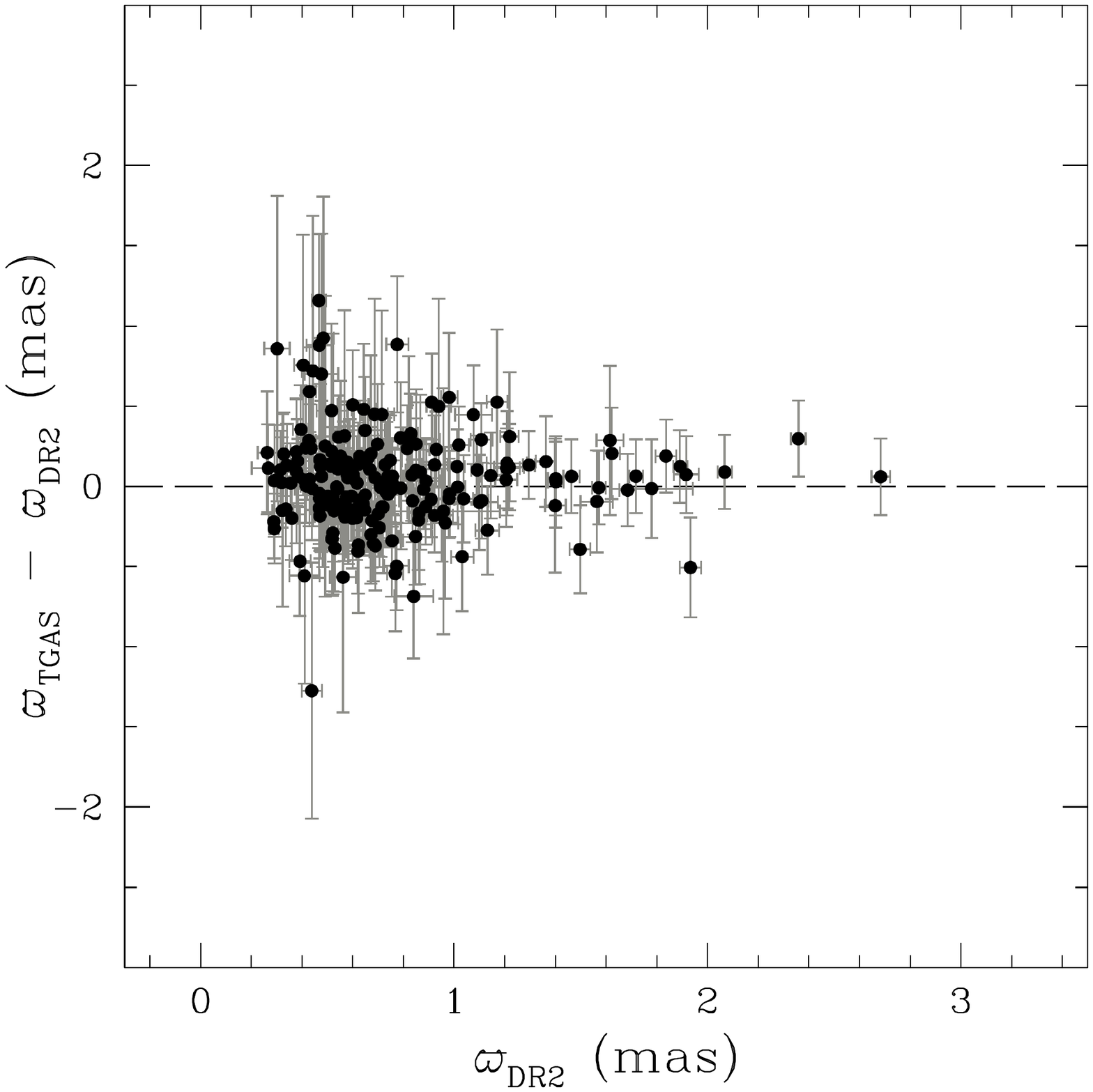}
  \caption{Comparison of the  TGAS and  DR2 parallaxes for 198 RRLs for which both parallax estimates are  available. For clarity, RR Lyr was omitted (see text for details).}
  \label{fig:comp_tgas_dr2}
\end{figure}

\begin{figure}
   \includegraphics[trim=30 170 30 100,width=1.05\linewidth]{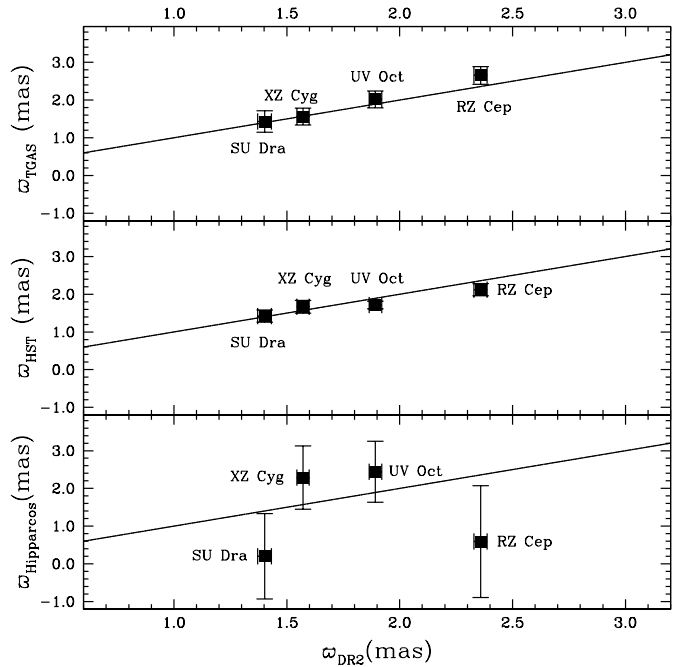}
  \caption{Comparison of the  DR2 and TGAS ({\it upper panel}), DR2 and HST ({\it middle panel}), and DR2 and Hipparcos ({\it lower panel}) parallaxes for RRLs studied by \citet{Benedict2011} with the {\it HST} (see also Table~\ref{tab:varhst}). RR Lyr was omitted (see text for the details). The solid lines represent the bisectors.}
  \label{fig:comp_parallax_all}
\end{figure}

\subsection{Comparison of the $PL$ relations\label{sec:pk}}
\begin{figure}
    \includegraphics[trim=30 180 30 100,width=1.05\linewidth]{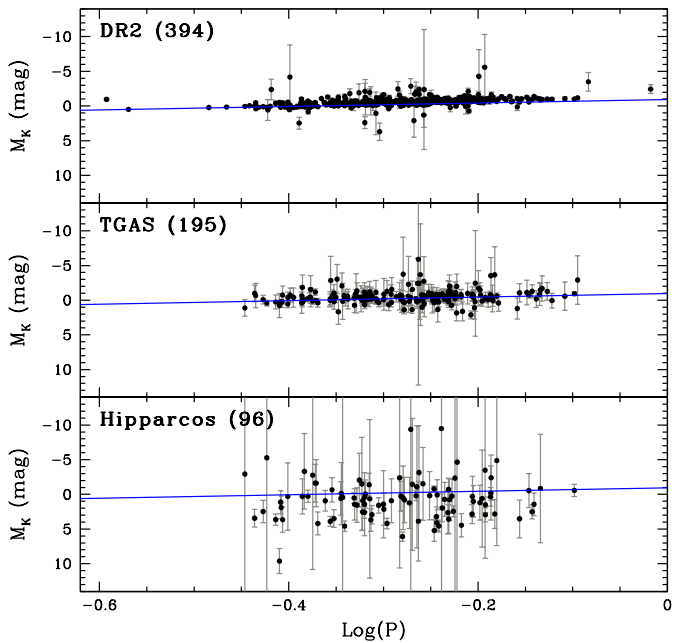}
  \caption{RRL  $PM_{K_{\rm s}}$ distributions obtained  inferring  the absolute magnitudes in the $K_{\rm s}$-band by direct transformation of the {\gaia} DR2 ({\it upper panel}), TGAS ({\it middle panel}) and {\it Hipparcos} ({\it lower panel}) parallaxes  using Eq.~\ref{eq:abs_mag}. Blue lines represent the $PM_K$ relation (eq. 14 from \citealt{Muraveva2015}). The period is measured in days.}
  \label{fig:comp_mk}
\end{figure}

\citet{Clementini2017} in their figure~23 show the impressive improvement of $PM_{K_{\rm s}}$ relation  of RRLs 
when going from the  Hipparcos to the TGAS parallaxes. In this section we extend the comparison 
to the DR2 parallaxes. 
To transform the trigonometric parallaxes to absolute magnitudes we used the canonical relation:

\begin{equation}\label{eq:abs_mag}
M = m_0 + 5\log \varpi  -10, 
\end{equation}
that links the star absolute magnitude $M$ and its  de-reddened apparent  magnitude $m_0$ to the star parallax in mas: $\varpi$. Eq.~\ref{eq:abs_mag} 
holds only  for  true  (hence with formally zero or negligible uncertainties) absolute  magnitudes,  apparent  magnitudes  and   parallaxes.
However, the direct transformation of parallaxes to the absolute magnitudes adopting Eq.~\ref{eq:abs_mag} cannot be used for measured values that have non negligible uncertainties (\citealt{Clementini2017}, \citealt{Luri2018}).
This is because the direct inversion of the measured parallaxes to estimate the distance is well behaved in the limit of negligible uncertainties, but degrades quickly as the fractional uncertainty of the parallax grows, resulting in estimates with large biases and variances.  
 Furthermore,  negative parallaxes cannot be transformed into absolute magnitudes, thus an additional sample selection bias is introduced, since objects with the negative parallaxes must be removed from the sample. The Bayesian fitting approach we apply in the present study allows us to avoid these issues as it is fully discussed in Section~\ref{sec:met}. However, only for visualisation purposes  in this section we transformed the Hipparcos, TGAS and DR2 parallaxes in the corresponding absolute $M_{K_{\rm s}}$ magnitudes using Eq.~\ref{eq:abs_mag}. This transformation is possible only for  394, 195 and 96 RRLs in our sample, for which positive parallaxes in the DR2, TGAS and Hipparcos  catalogues, respectively, are available. 
The corresponding $PM_{K_{\rm s}}$ distributions are shown in Fig.~\ref{fig:comp_mk}, that were obtained by correcting the apparent $K_{\rm s}$ mean magnitudes for extinction and after ``fundamentalizing" the RRc stars by adding 0.127 to the logarithm of the period. 
The $PM_{K_{\rm s}}$ distribution in the upper panel of Fig.~\ref{fig:comp_mk} shows the improvement of the DR2 parallaxes with respect to the TGAS 
(middle panel) and Hipparcos (lower panel) measurements. To guide the eye we plot as blue lines the $PM_K$ relation provided in eq.~14 by \citet{Muraveva2015}:
\begin{equation}
M_K = -2.53\log(P) - 0.95.\label{eq:pk_muraveva}
\end{equation}

The bottom panel of Fig.~\ref{fig:comp_mk} shows that the 96 RRLs with Hipparcos parallaxes are systematically shifted towards fainter absolute magnitudes. This is because by removing sources with negative parallaxes we are removing, preferentially,  RRLs at larger distances, of which Hipparcos  could measure only the brightest. Those distant bright RRLs typically will have small true parallaxes, close  to zero or even negative,  owing to the  larger uncertainties particularly in Hipparcos. Hence, the net effect of removing  RRLs with negative parallaxes in the Hipparcos sample is to bias the remaining  sample towards fainter absolute magnitudes as it is clearly seen in the bottom panel of Fig.~\ref{fig:comp_mk}.

\subsection{Comparison with Baade-Wesselink studies}\label{sec:bw}

\begin{figure}
\centering
\includegraphics[trim=30 170 30 100 clip, width=\linewidth]{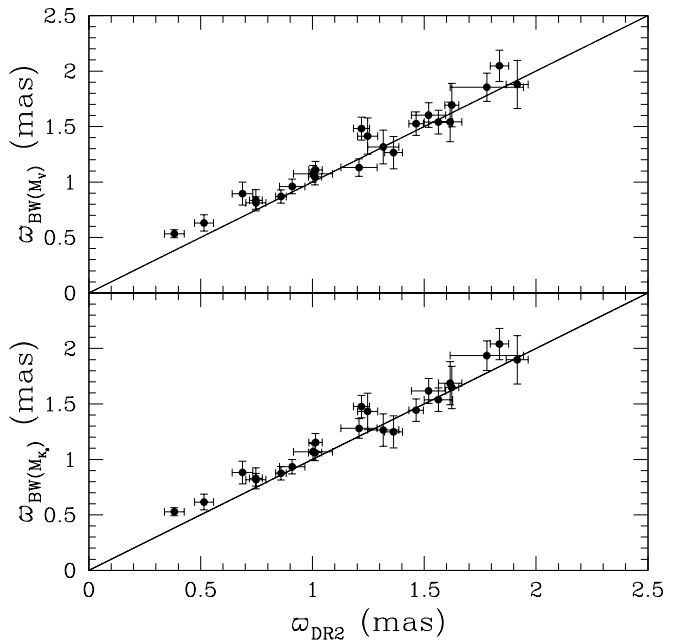}
\caption[]{Comparison of the  photometric parallaxes inferred from  $M_V$ ({\it upper panel}) and $M_{K}$ ({\it bottom panel}) absolute magnitudes  estimated
via  BW method and the corresponding DR2 parallaxes for 23 MW RRLs for which both parallax estimates are available. The solid lines represent the bisectors.}
 \label{fig:MkBWgaia}
\end{figure}

In this section we compare the DR2 trigonometric parallaxes with photometric parallaxes inferred from the Baade-Wesselink (BW)  technique,  which  are available for some of the RRLs in our sample. \citet{Muraveva2015} summarise in their table~2  the absolute visual  ($M_{V}$)  and $K$-band ($M_{K}$) magnitudes obtained  from the BW studies  (\citealt{Cacciari1992}, \citealt{Skillen1993}, \citealt{Fernley1998} and references therein) for 23 MW RRLs. The BW absolute magnitudes were revised assuming the value 1.38 \citep{Fernley1994} for the p factor used to transform the observed radial velocity to true pulsation velocity  and averaging multiple determinations for individual stars. All 23 RRL variables with absolute magnitudes estimated via BW technique have a counterpart in our sample of 401 RRLs. 
 The comparison of the photometric parallaxes inferred from the  BW $M_V$ and $M_{K}$  absolute magnitudes and the corresponding {\gaia} DR2
parallaxes for  these 23 RRLs  is shown in the upper and bottom panels of Fig.\ref{fig:MkBWgaia}, respectively.  
A weighted least squares fit of the relations  $\varpi_{M_{V(BW)}} =\alpha \varpi_{\rm DR2}$ 
and $\varpi_{M_{K(BW)}} =\alpha \varpi_{\rm DR2}$ returns the same slope value of 1.06,  which is  close to the bisector slope
$\alpha=1$. 
Even though the  DR2 parallaxes of these 23 RRLs are generally in good agreement with the photometric parallaxes obtained in the BW studies, we notice that there is a systematic difference between the two sets of parallaxes. Specifically, the mean non-weighted differences  $\Delta \varpi_0 = \varpi_{\rm DR2} - \varpi_{M_{V(BW)}}$ and $\Delta \varpi_0 = \varpi_{\rm DR2} - \varpi_{M_{K(BW)}}$ are both equal to  $-0.07$~mas. That is, the {\gaia} DR2 parallaxes for these 23 RRLs seem to be generally smaller than the photometric parallaxes inferred from the BW studies.  However, in the parallax offset estimate we used the direct transformation of the absolute magnitudes to parallaxes and assumed symmetric Gaussian uncertainties for the sake of simplicity.  Moreover, this offset is based on a relatively small number of close stars and depends on the specific value adopted for the p factor. We perform a more careful analysis of the potential {\it Gaia} DR2 parallax offset for RRLs and further discuss this topic in Section~\ref{sec:zp}.

\section{Method}\label{sec:met}

\subsection{Description of the Bayesian approach}\label{sec:bayes}  

In \cite{Delgado2018} we presented a Bayesian hierarchical method to
infer the $PM$ and $PMZ$ relationships. The hierarchical models were
validated with semi-synthetic data and applied to the sample of 200
RRLs described in \cite{Clementini2017}. Simplified versions of these
models were also used in \cite{Clementini2017}.  A full description of
these models is beyond the scope of this manuscript and we recommend
the interested reader to consult \cite{Delgado2018} for a more
in-depth description of them. In what follows, we summarize what we
consider are the minimum details about the hierarchical Bayesian
methodology and our models necessary to understand the present paper
as a self-contained study.

The core of the hierarchical Bayesian methodology consists of
partitioning the parameter space associated to the problem into
several hierarchical levels of statistical variability. Once this
partition is done, the modelling process assigns probabilistic
conditional dependency relationships between parameters at the same or
different levels of the hierarchy.  The construction of the model is
finished when one is able to express a joint probability function of
all the parameters of the model which factorizes as a product of
conditional probability distributions. This factorization defines a
Bayesian network \citep{Pearl1988,Lauritzen1996} consisting of a
Directed Acyclic Graph (DAG) in which nodes encode model parameters
and directed links represent conditional probability dependence
relationships.

All models presented in \cite{Delgado2018} have three-level
hierarchies. In all of them we partition the parameter space into
measurements ($\mathcal{D}$), true astrophysical parameters ($\Theta$)
and hyperparameters ($\Phi$).  In this case, the joint probability
distribution associated to our models is given by

\begin{equation}
	p\left(\mathcal{D},\mathbf{\Theta},\Phi\right)=p\left(\mathcal{D}
	\mid\mathbf{\Theta}\right)\cdot p\left(\mathbf{\Theta}\mid\Phi\right)\cdot p\left(\Phi\right)\,,
	\label{eq:joint-pdf-1}
\end{equation}

\noindent where $p\left(\mathcal{D}\mid\Theta\right)$ is the
conditional distribution of the data given the parameters (the so
called \textit{likelihood}), $p\left(\mathbf{\Theta}\mid\Phi\right)$
is the \textit{prior} distribution of the true parameters given the
hyperparameters and $p\left(\Phi\right)$ is the unconditional
\textit{hyperprior} distribution of the hyperparameters. Bayesian
inference is based on Bayes' rule:

\begin{equation}
p\left(\mathbf{\Theta},\Phi\mid\mathcal{D}\right)\propto
p\left(\mathcal{D}\mid\mathbf{\Theta}\right)\cdot
p\left(\mathbf{\Theta}\mid\Phi\right)\cdot p\left(\Phi\right)\,,
\label{eq:joint-posterior-pdf}
\end{equation}

\noindent and its goal is to infer the marginal posterior distribution
$p\left(\Lambda\mid\mathcal{D}\right)$ of some subset [$\Lambda\subseteq\left(\mathbf{\Theta},\Phi\right)$] of 
the parameters of interest. For the statistical inference problem 
of calibrating the RRL fundamental relationships,  in this paper we primarily 
focus on inferring the parameters of such relationships. However, we also aim at getting insight on 
the true posterior distribution of {\it Gaia} DR2 parallaxes given their measured values. 

\begin{figure}
	\begin{center}
                 \includegraphics[scale=.3]{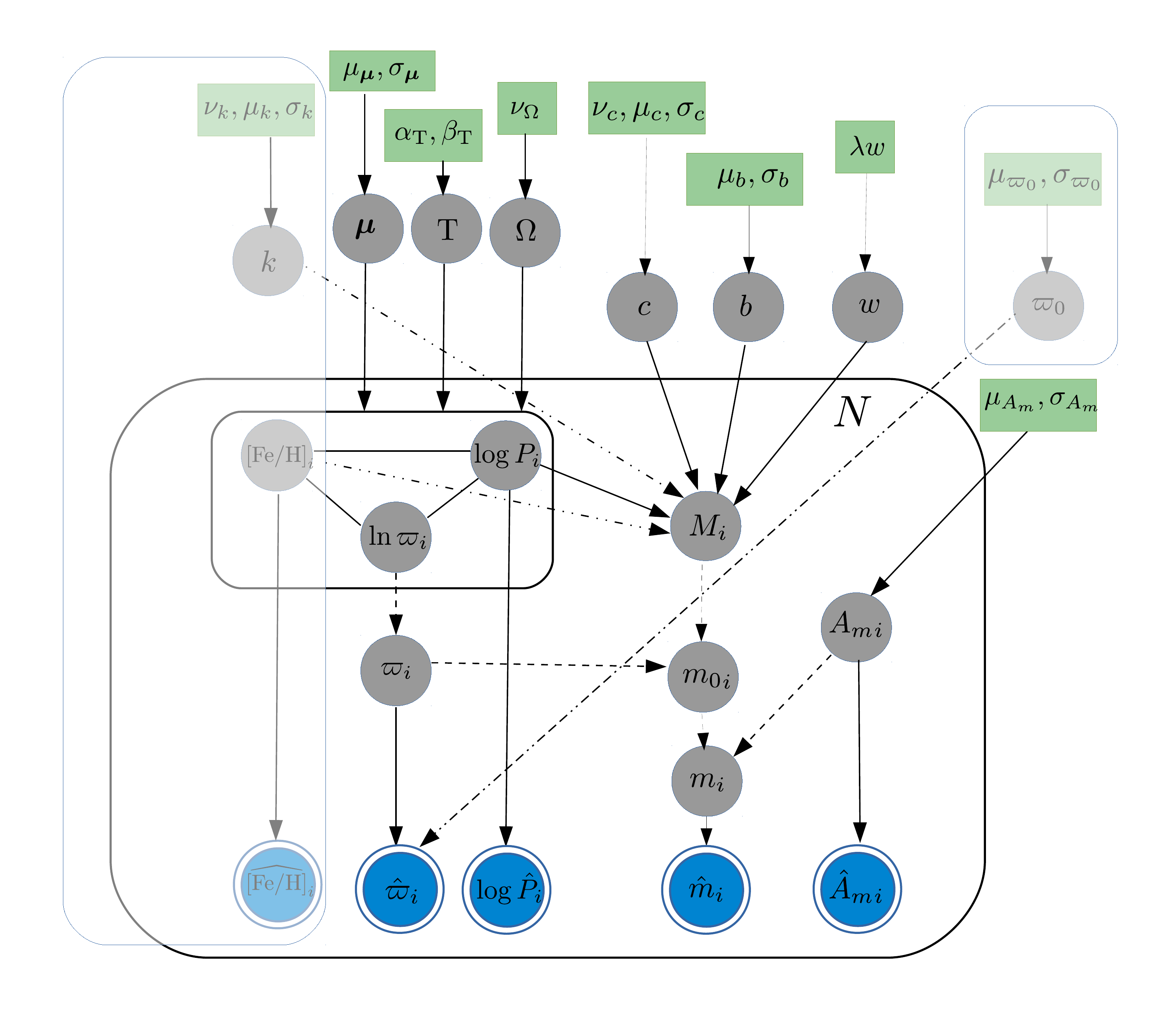}
                 \caption{Directed Acyclic Graph (DAG) that represents the
                  generic Hierarchical Bayesian Method (HBM) used to infer the coefficients of the
                  $PM$, $PMZ$ and optical absolute magnitude-metallicity ($MZ$) relations.}
		\label{DAG}
	\end{center}
\end{figure}

In Figure \ref{DAG} we present the generic DAG that encodes the
probabilistic relationships between the parameters of all the models
applied in this paper.  The graph is intended to represent several
models at once. For example, for a model devoted to infer a $PM$
relation, the nodes depicted with dimmer colours and their corresponding
arcs do not belong to the model and the reader should not consider
them.
The DAG of Fig.  \ref{DAG} shows the measurements at the bottom level
as blue nodes: decadic logarithm of periods $\log\hat{P}_{i}$,
apparent magnitudes $\hat{m}_i$, metallicities
$\widehat{\left[\mathrm{Fe/H}\right]}_{i}$, parallaxes
$\hat{\varpi}_i$ and extinctions $\hat{A}_{m_{i}}$ The subindex $i$
runs from 1 to the total number of stars $N$ in the sample. For each
measurement there is a corresponding true value in the DAG. True
values are represented as measurements but without the circumflex
accent (\verb!^!) and are depicted by grey nodes.  The model assumes
that all measurements are realizations from normal distributions
centred at the true (unknown) values and with standard deviations
given by the measurement uncertainties provided by each
catalogue. 
Note that true values and measurements are enclosed in a black
rectangle that represents sample replication for $i=1,2,...,N$ (plate
notation).
The model also
takes into account the effect of the interstellar absorption	
when generating the measured apparent magnitudes from the true ones. So, it
distinguishes between true (unabsorbed) apparent magnitudes ${m_0}_i$ and
true (absorbed) apparent magnitudes $m_{i}$ and deterministically computes the latter ones  
as $m_{i}={m_0}_i+A_{m_{i}}$, where  ${A_m}_i$ represents
the true absorption. This deterministic dependency of absorbed on
unabsorbed apparent magnitude and true extinction is represented by
dashed arrows in the DAG.
   
 
Once we have explained how the model manages the interstellar
absorption, we describe how it generates the true (unabsorbed)
apparent magnitudes ${m_0}_i$ from true parallaxes $\varpi_{i}$ and
absolute magnitudes $ M_{i}$. This is done by means of the
deterministic relationship, coming from Eq.~\ref{eq:abs_mag}:
 
 \begin{equation}
 	{m_0}_i  =  M_{i} -5\log(\varpi_{i})+10\,,
 	\label{eq:M-to-m0}
 \end{equation}
 
\noindent which is represented in the DAG by the dashed arcs going
from $\varpi_{i}$ and $ M_{i}$ to ${m_0}_i$.  The model also
contemplates the existence of a {\it Gaia} global parallax offset
$\varpi_{0}$ as suggested by \citet{Arenou2018}. This offset can be
inferred by the model itself or fixed to a predefined value.

The central core of the model is the submodel for the $PMZ$ relation
in which absolute magnitudes $ M_{i}$ are parameterized by

\begin{equation}
	M_{i}\sim\mathsf{t}_5\left(b+c\cdot\log
        P_{i}+k\cdot\left[\mathrm{Fe/H}\right]_{i},w\right)\,,
	\label{eq:PLZ}
\end{equation}

\noindent  where $\sim$ should be read as `is distributed as' and
  $\mathsf{t}_5$ represents a Student's t distribution with five degrees
  of freedom. The mean of this distribution is a linear model in three
  parameters: the intercept $b$, the slope $c$ for the period term (in
  decadic logarithmic scale), and the slope $k$ for the metallicity
  term, and its scale  $w$ represents the linear model
  intrinsic dispersion. This intrinsic dispersion aims at including
  all potential effects not accounted for explicitly in the model
  (e.g. evolutionary effects). The dependency of absolute magnitude on
  the linear model parameters, intrinsic dispersion and predictive
  variables is represented in the DAG by all the incoming arrows to
  the node $M_{i}$. A Student's t distribution was chosen in order to
  make the model robust against outliers such as RR~Lyr itself and a few
  other stars with potentially problematic DR2 parallax and parallax
  uncertainty values.

For the slopes $c$ and $k$ of the $PMZ$  
relationship of
Eq. \ref{eq:PLZ} we specify weakly informative Student's t-priors
with location parameter $\mu_{c}=\mu_{k}=0$, and for the intercept we
use a vague Gaussian prior centred at $\mu_{b}=0$. The intrinsic
dispersion of the $PMZ$ relation is given an exponential prior with
inverse scale $\lambda_w=1$. We use green rectangular nodes at the top
of the graph to denote all these fixed prior hyperparameters.

In \cite{Delgado2018} we reported the existence of a systematic
correlation between measured periods, metallicities and parallaxes in
our sample. This correlation occurred in the sense that a slight
decrease of the median period calculated in bins of parallax
corresponds to an increase of the median parallax and the median
metallicity.  We also demonstrated that without a proper modelling of
this systematic effect the inference carried out with the TGAS parallax 
measurements returned a severely underestimated slope in $\log(P)$  of the 
$PMZ$ relation. 
We also showed that for the typical DR2 parallax
uncertainties, the impact of the underestimation is severely 
reduced. In any case, our model assigns a joint prior to the true
values of $\log P_i$  and
$\varpi_{i}$ in order to prevent this bias. This prior is defined
as

\begin{equation}
\left(\log P_{i},\ln\varpi_{i}\right)
\sim\mathsf{MN}\left(\boldsymbol{\mu},\mathrm{T}\Omega\mathrm{T}\right)\,,
\label{eq:prior-PZPi}
\end{equation}

\noindent where $\mathsf{MN}$ represents a 2D Gaussian distribution
with mean vector $\boldsymbol{\mu}=(\mu_{P},\mu_{\omega})$,
diagonal matrix of standard deviations
$\mathrm{T}=\mathrm{diag}(\sigma_{P},\sigma_{\omega})$ and
correlation matrix $\Omega$. We parametrize each component of the mean
vector $\boldsymbol{\mu}$ by a weakly informative Gaussian prior 
  centred at 0 and with standard deviation
  $\mathrm{\sigma}_{\boldsymbol{\mu}}=5$. For each standard deviation
in $\mathrm{T}$ we assign a weakly informative Gamma distribution
prior with shape $\alpha_{\mathrm{T}}=0.1$ and inverse scale
$\beta_{\mathrm{T}}=0.1$.  For the correlation matrix $\Omega$ we
specify a LKJ prior \citep{LEWANDOWSKI20091989} with $\nu_{\Omega}=1$
degrees of freedom.
 To every true value of ${\left[\mathrm{Fe/H}\right]}_{i}$ our model
assigns a non-informative Gaussian prior which reflects our 
limited knowledge about the true distribution of metallicities, given
the heterogeneous provenance of metallicities in \citet{Dambis2013} sample.

\section{Characteristic relations for RRLs}\label{sec:relations}
\subsection{{\mz} relation}\label{sec:mz}

A vast and long-standing literature exists on the visual {\mz}
relation of RRLs (see e.g. \citealt{Clementini2003};
\citealt{Cacciari2003}; \citealt{Dambis2013} and references
therein). The relation is generally assumed to have a linear form:
$M_V = \alpha[Fe/H]+\beta$, with literature values for the slope
$\alpha$ of the metallicity term ranging from $0.30-0.37$
\citep{Sandage1993,Feast1997} to 0.13 \citep{Fusi1996} and an often
adopted mild slope of $\alpha = 0.214\pm0.047$ mag/dex, as estimated
from a photometric and spectroscopic study of about a hundred RRLs in
the bar of the LMC (\citealt{Clementini2003};
\citealt{Gratton2004}). Those studies also showed the {\mz} relation
to be, in first approximation, linear and universal.  On the other
hand, theoretical studies (e.g. \citealt{Caputo2000b};
\citealt{Bono2003}; \citealt{Catelan2004}) suggest that the {\mz}
relation is not linear over the whole metallicity range spanned by the
MW RRLs (almost 3 dex for MW field variables). Indeed, theoretical
studies (e.g. \citealt{Caputo2000b}; \citealt{Bono2003}) state that
$\alpha$ presents
a relatively mild value of $0.17 \pm 0.069$ mag/dex for RRLs with
${\rm [Fe/H]} \le -1.6$~dex and becomes significantly steaper, $0.359
\pm 0.027$ mag/dex, for RRLs with ${\rm [Fe/H]} > -1.6$~dex.

In this paper we study the {\mz} relation of RRLs applying the
Bayesian approach described in Section~\ref{sec:met} to 381 RRLs, out
of our sample of 401, for which all needed information (apparent $V$
mean magnitudes, extinction, metal abundances) are available
from \citet{Dambis2013} and trigonometric parallaxes are available in
{\it Gaia} DR2.  Slope, zero-point and intrinsic dispersion obtained
for the {\mz} relation with our approach are summarised in the first
row of Table~\ref{tab:offset-comp-linear-LZ-DR2}.  The
  metallicity slope we obtain, $\alpha = 0.40 \pm 0.03$ mag/dex,
  implies a much stronger dependence of the RRL absolute $V$ magnitude
  on metallicity than ever reported previously in the literature.  In
Section~\ref{sec:zp} we argue that this steep slope may be due to an
offset affecting the DR2 parallaxes. However, we first check here
whether the high metallicity dependence is not caused by a flaw in our
Bayesian procedure.

\begin{table*}
\caption{{\mz}  relation of RRLs in linear form obtained with our Bayesian approach when a {\it Gaia} DR2 parallax zero-point offset, $\Delta \varpi_0$, is fixed in the model: (1) number of stars used in the fit; (2) and (3) slope and zero-point of the relation; (4) intrinsic dispersion; (5)  adopted value for the {\it Gaia} DR2 parallax zero-point offset.} 
	\begin{center}
		\begin{tabular}{cccccr}
		\hline
\hline
			&	No. stars & $\alpha$  &  $\beta$ & $w$ & $\Delta \varpi_0$\\
			\hline
			&	$381$ & ${0.40}_{-0.03}^{+0.03}$ &  ${1.01}_{-0.04}^{+0.04}$ & ${0.17}_{-0.01}^{+0.01}$ & $0.00$ \\
			&	$381$ & ${0.37}_{-0.03}^{+0.03}$ &  ${1.09}_{-0.04}^{+0.04}$ & ${0.15}_{-0.01}^{+0.01}$ & $-0.03$ \\
			&	$381$ & ${0.33}_{-0.02}^{+0.03}$ &  ${1.19}_{-0.04}^{+0.04}$ & ${0.15}_{-0.01}^{+0.01}$ & $-0.07$ \\
			\hline
			&	$23$ & ${0.27}_{-0.06}^{+0.06}$ &  ${0.94}_{-0.08}^{+0.08}$ & ${0.14}_{-0.03}^{+0.04}$ & $0.00$ \\
			&	$23$ & ${0.27}_{-0.06}^{+0.05}$ &  ${0.99}_{-0.08}^{+0.08}$ & ${0.13}_{-0.03}^{+0.04}$ & $-0.03$ \\
			&	$23$ & ${0.26}_{-0.05}^{+0.05}$ &  ${1.06}_{-0.07}^{+0.07}$ & ${0.13}_{-0.03}^{+0.04}$ & $-0.07$ \\
			\hline
		\end{tabular}
		\label{tab:offset-comp-linear-LZ-DR2}
	\end{center}
\end{table*}
To this end  we 
carried out simulations with semi-synthetic data. The semi-synthetic data were created according to the following recipe:
\begin{itemize}

\item we generated true metallicities drawn from Gaussian
distributions (one for each star) centred at the measured values and
with standard deviations given by the uncertainties described in
Section \ref{sec:data};

\item absolute magnitudes $M_V$ were derived from the true metallicities
generated above using the {\mz} relationship by \citet{Federici2012}:
$M_V=0.25{\rm [Fe/H]} + 0.89$; 

\item true de-reddened apparent magnitudes were generated
from Gaussian distributions centred at the observed values and with
standard deviations given by the measurement uncertainties corrected
for  $V$-band extinctions generated likewise centred at the values
in  \citet{Dambis2013};

\item finally, the true parallaxes were derived from the true absolute
and apparent magnitudes, and the measured parallaxes were drawn from
Gaussian distributions centred at the true values with standard
deviations given by the {\it Gaia} DR2 parallax uncertainty of each star.

\end{itemize}
Our  aim is to evaluate the capability of our methodology to recover unbiased
estimates of the true parameters, in particular of the relationship
slope. We infer posterior distributions of the model parameters for 10
realisations of the semi-synthetic data. The inferred slopes range
between 0.238 and 0.274 mag/dex with a mean of 0.259 mag/dex (the true value used in
the generation of the data being 0.25) with a constant credible
interval of $\pm 0.05$. This represents a mild overestimation of
the true slope which could be due to the imperfection of the assumed priors and the relatively
small number of simulations. 
We have also applied models with a prior for the parallax that is
independent of the period. As discussed in \cite{Delgado2018},
this neglects the existence of a correlation between parallax and period in our sample and, as
expected, results in slopes that are systematically biased (yielding
an average slope from 10 random realisations of the semi-synthetic
data set of 0.31 mag/dex and a posterior standard deviation of 0.05). 
In summary, we find that  the Bayesian
model designed and applied to the data is not the cause for the slopes being higher than expected
according to the previous studies.

\begin{figure}
   \includegraphics[trim=0 80 0 70,width=\linewidth]{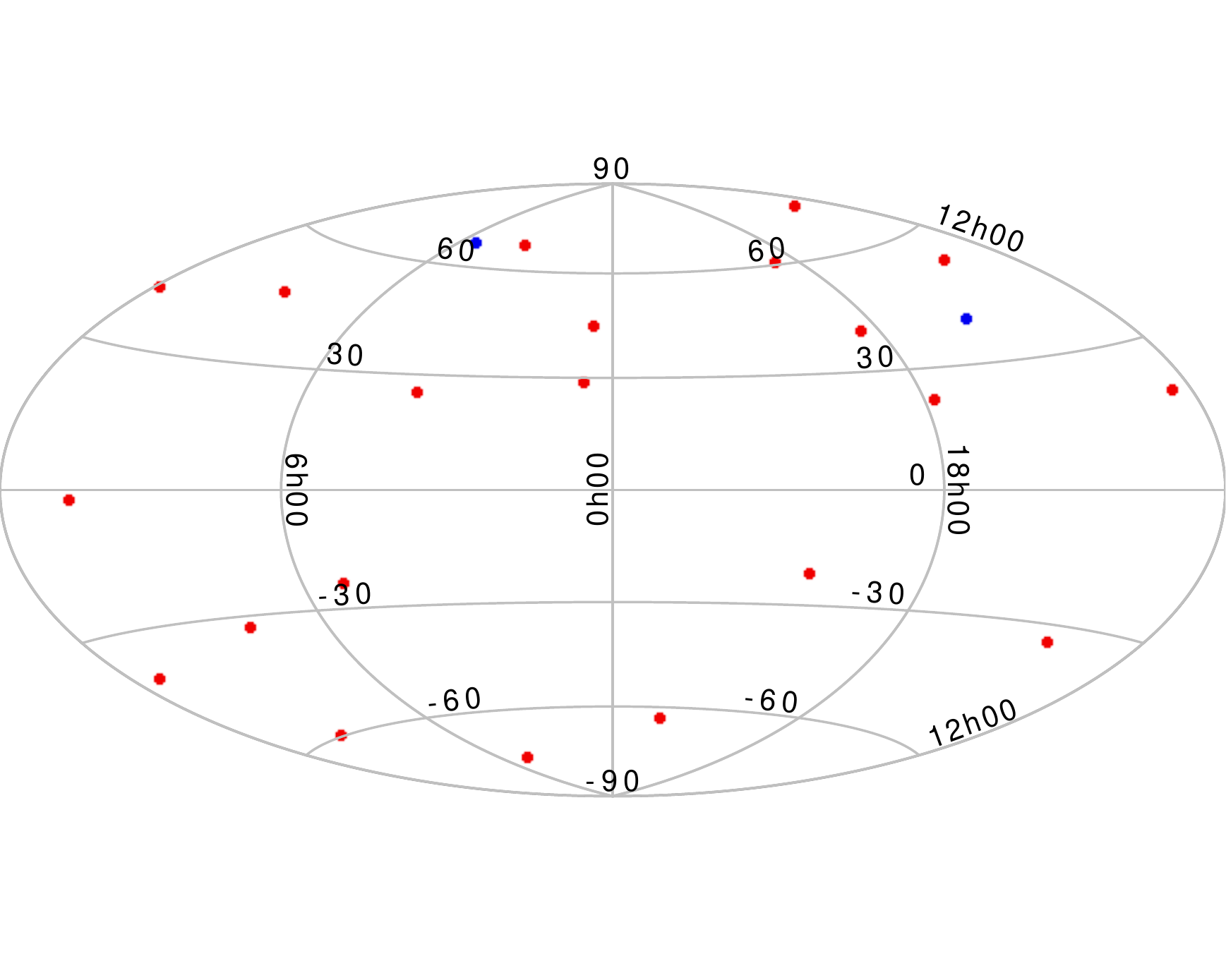}
  \caption{Sky distribution in  Galactic coordinates of 23 RRLs for which absolute magnitudes in the $V$ and $K$ bands have been estimated using the BW technique. Red and blue filled circles show RRab and RRc stars, respectively.}
  \label{fig:map23}
\end{figure}

\begin{figure}
    \includegraphics[trim=20 40 0 10,width=1.05\linewidth]{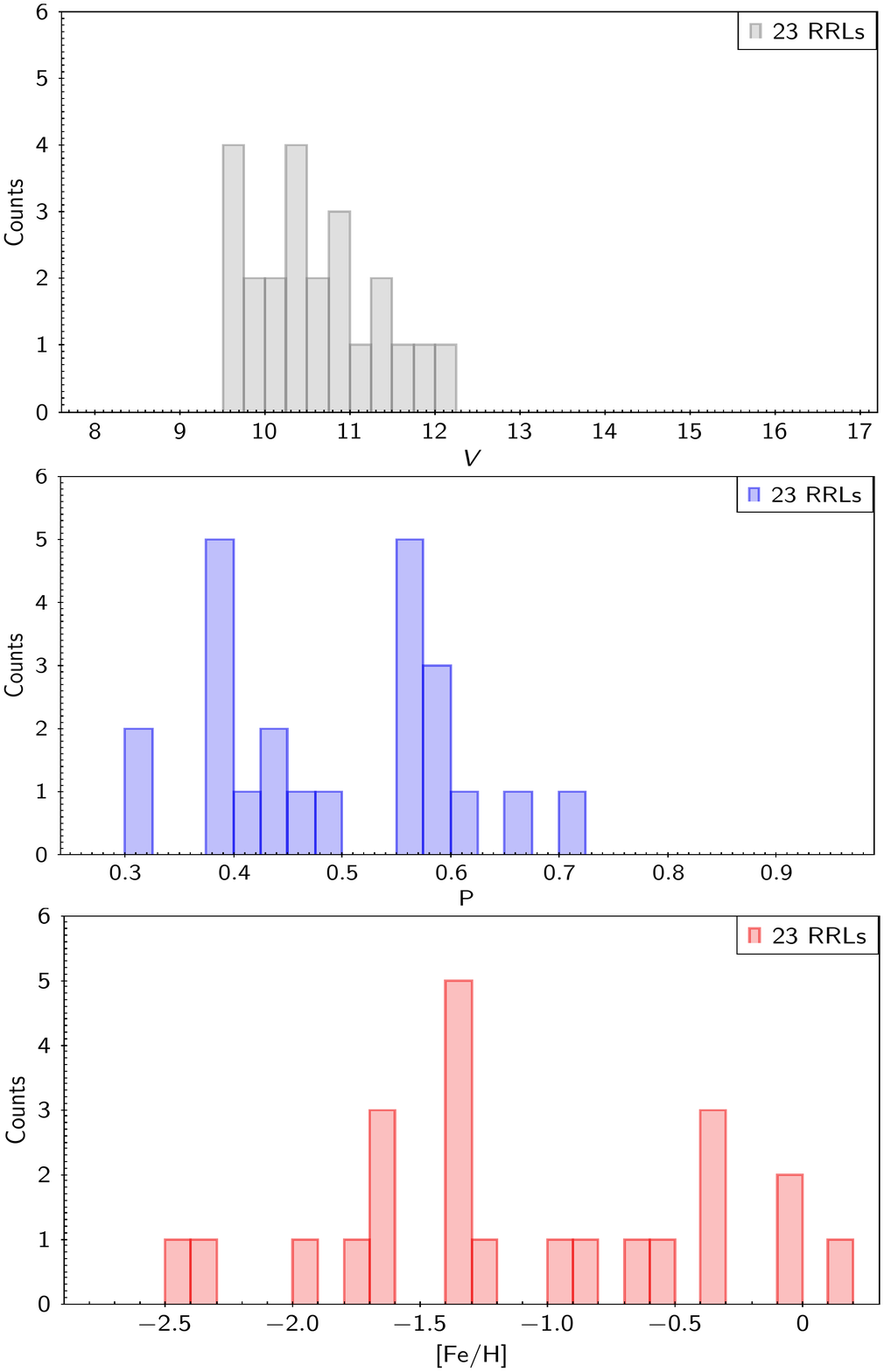}
  \caption{Distribution in apparent $V$ magnitude ({\it upper panel}), period ({\it middle panel}) and metallicity (as derived from high-resolution spectroscopy; {\it lower panel}) for the 23 RRLs 
  in Fig.~\ref{fig:map23}.}
  \label{fig:hist_23}
\end{figure}

The steeper slope found by our methodology can be compared with
alternative inference methods commonly used in the literature. We can compare the results with the
weighted least-squares estimate when the weights are inversely
proportional to the uncertainties in both the predicting and predicted
variables. We interpret that this is the methodology applied when it is
claimed in the literature that the fit was a weighted linear
least-squares model with uncertainties in both axes. Details about
how  uncertainties are combined are often missing in the literature. We will
assume that a simple quadratic addition was put in place. However,
this is far from optimal as it does not include a full forward model
of how the data were created and does not distinguish between
uncertainties in absolute magnitude and metallicity. The effects of
uncertainties in each of the axes are significantly different in size 
and nature,  hence we only include these results for the
sake of comparison. If we apply this weighted least-squares
approximation to the absolute magnitudes  naively computed by parallax
inversion (Eq.~\ref{eq:abs_mag}),  as a consequence, truncating the sample by removing negative
parallaxes,  we obtain a slope value  of $\alpha = 0.36\pm0.03$. 
Therefore, it seems 
there is evidence for a
steeper slope even from the less sophisticated method that does not
include intrinsic dispersion, does not deal correctly with the
uncertainties in the absolute magnitude and truncates the sample. 
We conclude that the high metallicity slope of the {\mz} relation is not caused by the selected inference method, and rather reflects 
the real distribution of the data. In this case,  selecting a sample with accurately homogenised  metallicity estimates becomes crucially important. Metallicity estimates used for our study are provided by  \citet{Dambis2013}, who in their turn  compiled  metal abundances obtained in a number of different studies and include metallicity  values inferred from  the Fourier analysis of the RRLs light curves or using high- or low-resolution spectroscopy. For the sake of homogeneity \citet{Dambis2013} transformed  the estimated metallicity values to the unique \citet{ZW}  metallicity scale. This transformation could cause an additional bias which is hard to account for. Furthermore, uncertainties in metallicities were not provided by \citet{Dambis2013}, hence, we assigned approximate values of uncertainties (see Section~\ref{sec:data}), that could also affect the results of our fit. In order to avoid all these issues we decided to use a sample of  23 MW RRLs studied by \citet{Muraveva2015}, for which homogeneous  metallicities and related  uncertainties based on abundance analysis of  high-resolution spectroscopy (\citealt{Clementini1995},  \citealt{Lambert1996}) are available. All 23 RRLs have counterparts in our sample of 401 RRLs. Their distribution on the sky is shown in Fig.~\ref{fig:map23}. Their apparent $V$ magnitudes and periods expectably span narrower ranges than the full sample of 401 RRLs, namely,  from 9.55 to 12.04~mag in apparent visual mean magnitude and from 0.31 to 0.71 days in period. Metal abundances for these 23 RRLs are in the range from $-2.5$ to +0.17~dex, which is comparable with the range of metallicities spanned by our 401 RRL full sample (see Section~\ref{sec:data}). The distributions in $V$ magnitude, period and metallicity of the 23 RRLs are shown in Fig.~\ref{fig:hist_23}, where for  ease of comparison we use the same scales for the abscissa axes as in Fig.~\ref{fig:hist_all}. Slope, zero-point and dispersion of the RRL {\mz} relation obtained by applying our Bayesian approach to the sample of 23 RRLs are summarised in the first row of the 
lower portion of Table~\ref{tab:offset-comp-linear-LZ-DR2}. The metallicity slope ($\alpha = 0.27 \pm 0.06$) is shallower than obtained from the whole sample of  381 RRLs, that may be the effect of 
using more accurate and homogeneous spectroscopic metallicities,  and yet is still steeper than found in the more recent literature, even though (marginal) agreement exists within the uncertainties. 
A clue to further investigate the metallicity dependence issue may be to study the {\mz} relation defined by large samples of RRLs in globular clusters for which the metallicity is well known from high resolution spectroscopic studies. This is addressed  in a following paper (Garofalo et al., in preparation).  Meanwhile, in Section~\ref{sec:zp} we examine  whether the existence of a zero-point offset in the {\it Gaia} DR2 parallaxes  (\citealt{Arenou2018}) might contribute to produce the high slope we find for the RRL  {\mz} relation.

\subsection{{\it \textbf{Gaia}} DR2 parallax offset}\label{sec:zp}
The {\gaia} DR2 parallaxes  are known to be affected by an overall zero-point whose extent varies depending on the sample used to infer its value and, typically,  is of the order of $-0.03$~mas, as inferred  by comparison with QSO parallaxes (\citealt{Arenou2018}). After {\gaia} DR2  a number of studies have appeared in the literature (\citealt{Riess2018}; \citealt{Zinn2018}; \citealt{Stassun2018}) all suggesting that 
{\it Gaia}  provides smaller parallaxes, hence, larger distance estimates,   than derived from other  independent techniques, that is,  $\Delta \varpi_0 = \varpi_{\rm DR2} - \varpi_{indep.}$ is negative, but by how much varies from one study to the other. For instance, \citet{Riess2018} estimated a zero-point offset for {\gaia} DR2 parallaxes of $-0.046$~mas, combining {\it HST} photometry and {\it Gaia} DR2 parallaxes for a sample of 50 Cepheids. 
In the paper describing the final validation of all data products published in {\it Gaia} DR2,  \citet{Arenou2018} 
estimate an offset $\Delta \varpi_0 = -0.056\pm0.005$~mas for RRLs in the {\it Gaia} DR2 catalogue and $-0.033 \pm 0.009$~mas for a sample of RRLs in the GCVS. 
In the following we investigate whether the zero-point offset of the {\it Gaia} DR2 parallaxes for RRLs can affect the slope and zero-point inferred for their {\mz} relation.
We remind the
reader that the non-linear relationship between parallax and absolute
magnitude (Eq.~\ref{eq:abs_mag}) implies that a given parallax offset does not affect all absolute magnitudes equally. Just for illustration purposes, a
parallax offset of $-0.056$~mas  as suggested by \citet{Arenou2018} for the RRLs would result in a change of the distance modulus equal to 0.2~mag at 1 kpc, while at 7 kpc it would amount to 1.1~mag. We are located in a relatively metal-rich area of the MW, while farther RRLs in our sample belong to the halo and likely are more metal-poor. Thus, the negative zero-point offset in the DR2 parallaxes will make farther/metal-poor RRLs to appear intrinsically brighter, hence causing an overestimation of the {\mz} relation slope. 
Therefore, our discussion of the results must incorporate the effect of potential parallax offsets.
 The upper portion of Table \ref{tab:offset-comp-linear-LZ-DR2} lists inference results 
  for a series of linear {\mz} 
models characterised by different parallax offsets, namely, 
models without offset, with a
global offset of $-0.03$~mas (\citealt{Arenou2018}) and with an offset  of $-0.07$~mas, as suggested by the
comparison with the  absolute magnitudes derived from the BW
studies (Section~\ref{sec:bw}) and used here only for demonstration reasons. Results of this test show that it is of crucial importance to take into account a potential parallax offset when studying 
the RRL {\mz} relation because the slope of the relation  
varies  with the offset and decreases systematically with increasing  the value adopted for the parallax offset, 
from $\alpha=0.40$ (no offset),  to $\alpha=0.33$ for an offset of $-0.07$~mas. 
We applied the same model to the 23 MW RRLs with homogeneous
metallicity estimates based on abundance analysis of high-resolution spectra
(see Section~\ref{sec:mz}), assuming no parallax offset and the offsets of $-0.03$~mas and $-0.07$~mas. The resulting relations are shown in the lower portion of Table~\ref{tab:offset-comp-linear-LZ-DR2}. The slope of the {\mz} relation varies from $0.27$ (no offset) to $0.26$ (offset of $-0.07$ mas), showing that for the sample of 23 RRLs,  which have both  distances  and range of distances much smaller than for the full sample,  the impact of a potential parallax offset on the slope of the {\mz} relation is greatly reduced.

\begin{figure}
	\begin{center}
                \includegraphics[trim=20 50 40 10, width=\linewidth]{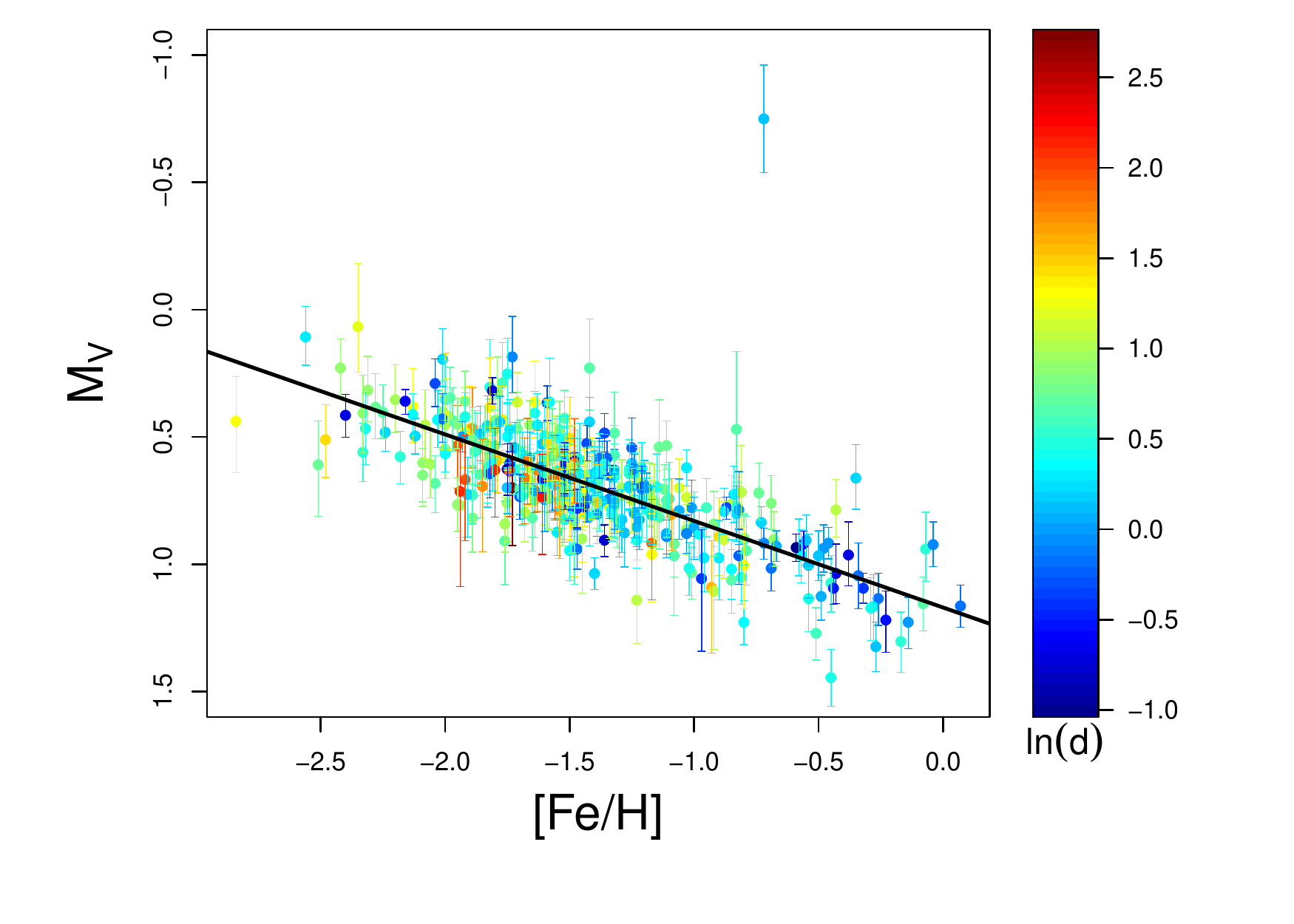}
		\caption{{\mz} relation defined by the  381 RRLs  in our sample,
			whose absolute $M_V$ magnitudes were inferred from the model
			described in Section \ref{sec:met}. The solid line represents the linear fit. Its slope and zero-point 
			are summarised in the first portion of Table \ref{tab:fittings-DR2} (first row). The colour scale 
			encodes the (natural) logarithm of the inferred (true) distance measured in units of kpc.}
		\label{381-MV-Z-linear-fit}
	\end{center}
\end{figure}

To compute all the relationships presented in this paper we use the model that includes the potential parallax offset as a parameter (Section~\ref{sec:met}). 
We fitted the {\mz} relation defined by the whole sample of RRLs inferring simultaneously the relation parameters (slope and zero-point) and the parallax offset. Corresponding results 
are summarised in the first panel of Table~\ref{tab:fittings-DR2} (first row). From our sample of 381 RRLs we obtain a mean posterior metallicity slope of 0.34 for a mean posterior offset of $-0.062$~mas. The resulting {\mz} fit is shown  in Fig.~\ref{381-MV-Z-linear-fit}, where colours encode the (natural) logarithm of the inferred distance. 

\begin{figure}
	\begin{center}
                \includegraphics[trim=20 50 40 10, width=\linewidth]{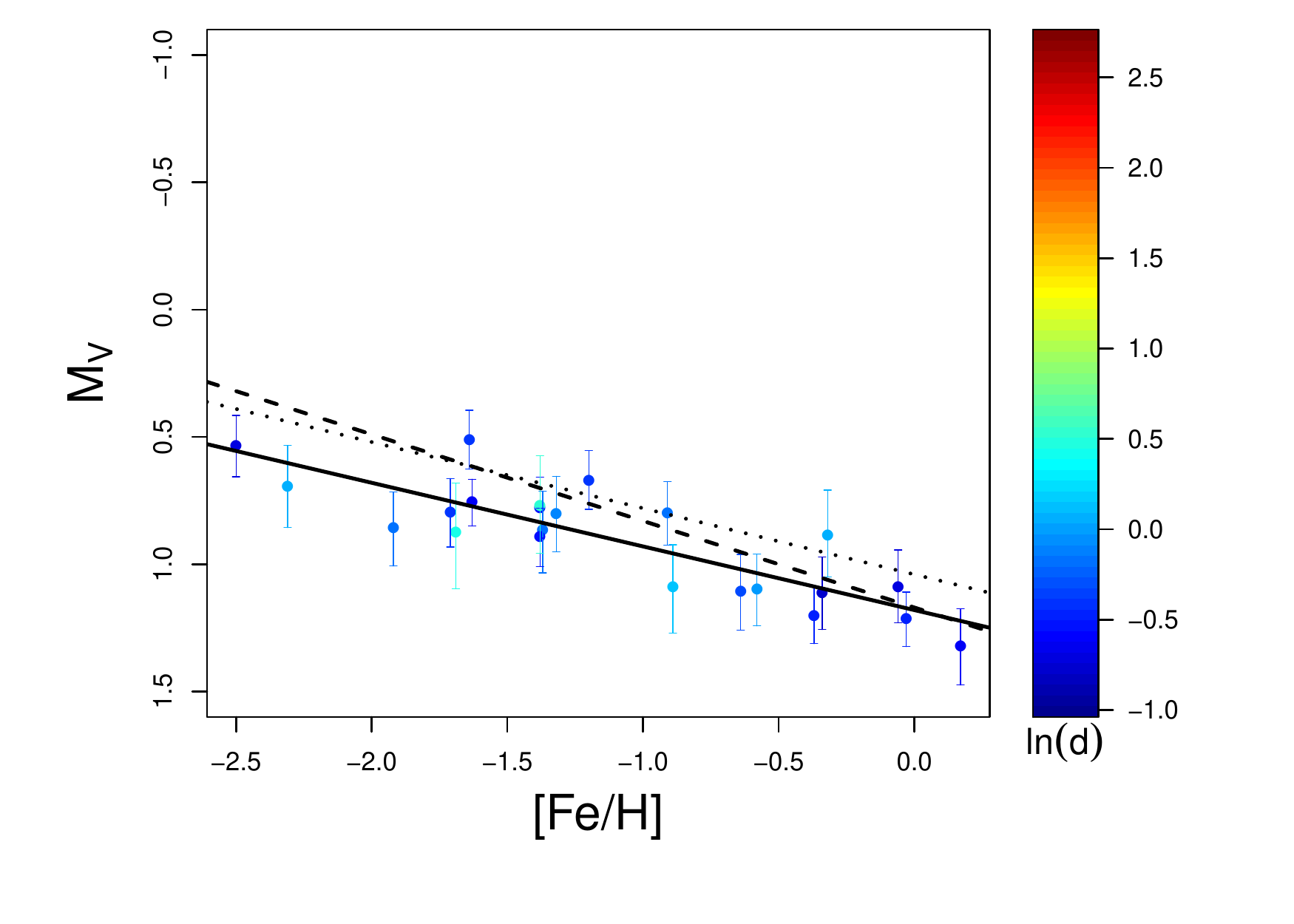}
		\caption{{\mz} relation defined by the 23 RRLs  with metallicity from high resolution spectroscopy,
			whose absolute $M_V$ magnitudes were inferred from the model
			described in Section \ref{sec:met}. The solid line represents the linear fit. Its slope and zero-point are 
			summarised in the second  portion of Table \ref{tab:fittings-DR2}. For comparison, the dashed line represents the 
			relationship inferred from the $381$ RRLs shown in the first portion of the same Table.
			The dotted line represents the relationship inferred by a model for the 23 RRLs with a parallax offset
			 fixed to $-0.057$ mas and shown in the third portion of  Table~\ref{tab:fittings-DR2}. The colour 
			 scale encodes the (natural) logarithm of the inferred (true) distance measured in units of kpc.}		
		\label{23-MV-Z-linear-fit}
	\end{center}
\end{figure}

We applied the same model to the 23 MW RRLs with homogeneous
metallicity estimates. The resulting {\mz} relation is shown in the second panel of Table~\ref{tab:fittings-DR2}.  In the case of the 23 MW RRLs, the reduced number of 
sources
and the smaller range of distances (visible from the colour scale in Fig.~\ref{23-MV-Z-linear-fit}) 
do not constrain the value of the
offset. The first row of the second portion of Table~\ref{tab:fittings-DR2} summarises the posterior distribution for the
offset with a mean of $-0.142^{+0.06}_{-0.06}$ which seems implausible. This
translates directly into a much smaller metallicity slope of 0.25 $\pm$ 0.05 mag/dex compared to the
one inferred for the 381 stars sample. If we remove the determination of the parallax offset as a parameter of the model and, instead, adopt a constant value for the offset  of $-0.057$ mas, 
which corresponds to the average of the offsets obtained from fitting the 
 linear {\mz} relation ($-$0.062 mas; Section~\ref{sec:mz}), and the  {\pkz}, {\pwz} relations ($-$0.054 and $-$0.056 mas, respectively; Section~\ref{sec:pkz}) to the full sample of RRLs, we obtain a slope of $0.26\pm 0.05$ for the  sample of 23 RRLs. The resulting {\mz} relation is shown in the third portion of Table~\ref{tab:fittings-DR2}.

 To conclude even though using  the reduced sample of 23 RRLs has the advantage of (i) a smaller effect of the parallax offset as the 23 RRLs are nearby stars; (ii) a more accurate estimation of metallicity based on high-resolution spectroscopy, we must stress that selection effects can potentially be stronger, as only nearby bright RRLs are characterised by
high enough signal-to-noise ratios to be analysed with high-resolution spectroscopy, hence, biasing the results.

\begin{figure}
	\begin{center}
		\includegraphics[trim=20 50 40 10, width=\linewidth]{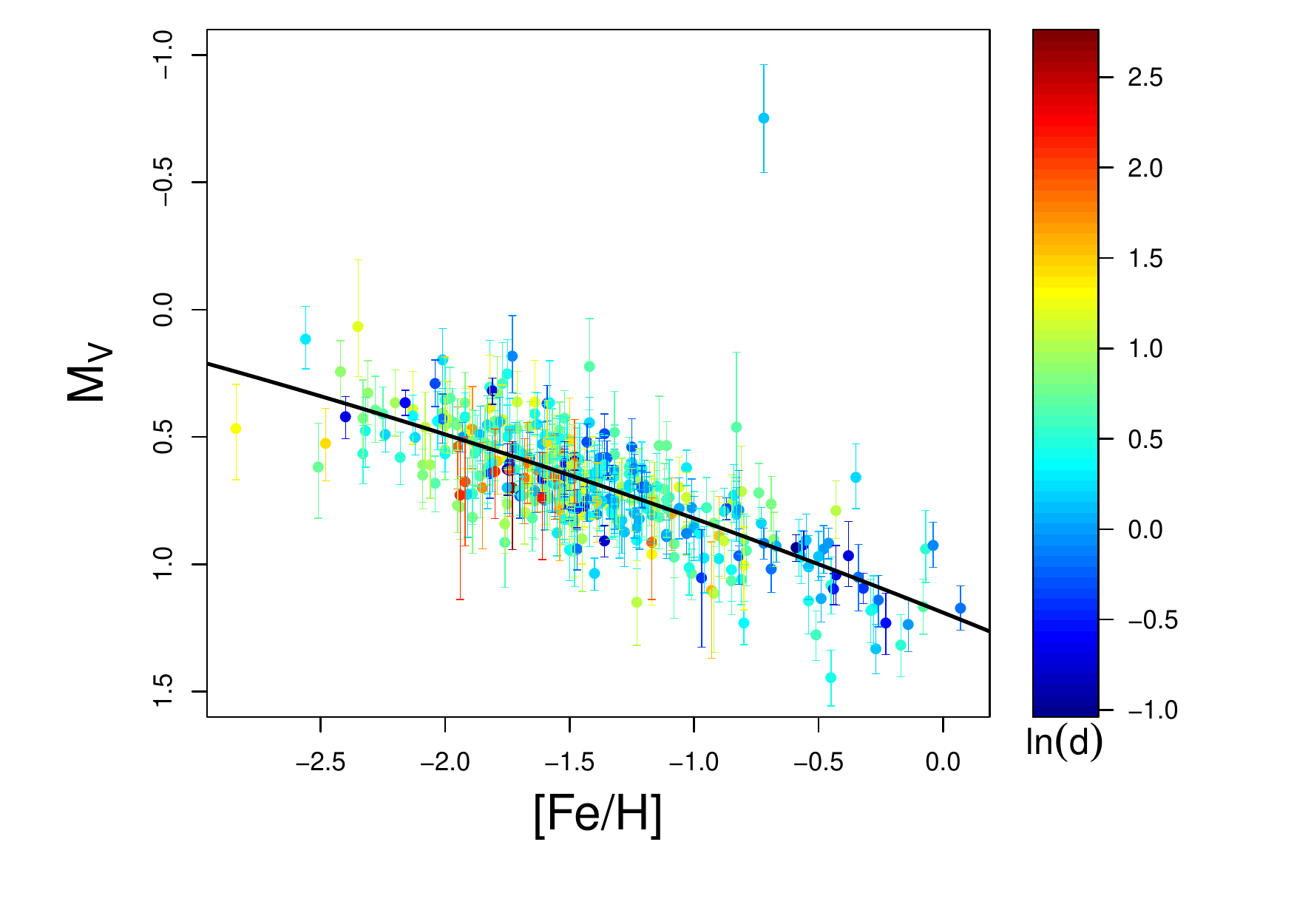}
		\caption{Same as in Fig.~\ref{381-MV-Z-linear-fit},  but with the solid line representing the  quadratic {\mz} relation (second row of the first portion of Table \ref{tab:fittings-DR2}).}
		\label{381-MV-Z-quadratic-fit}
	\end{center}
\end{figure}

\begin{figure}
	\begin{center}
	        \includegraphics[trim=20 50 40 10, width=\linewidth]{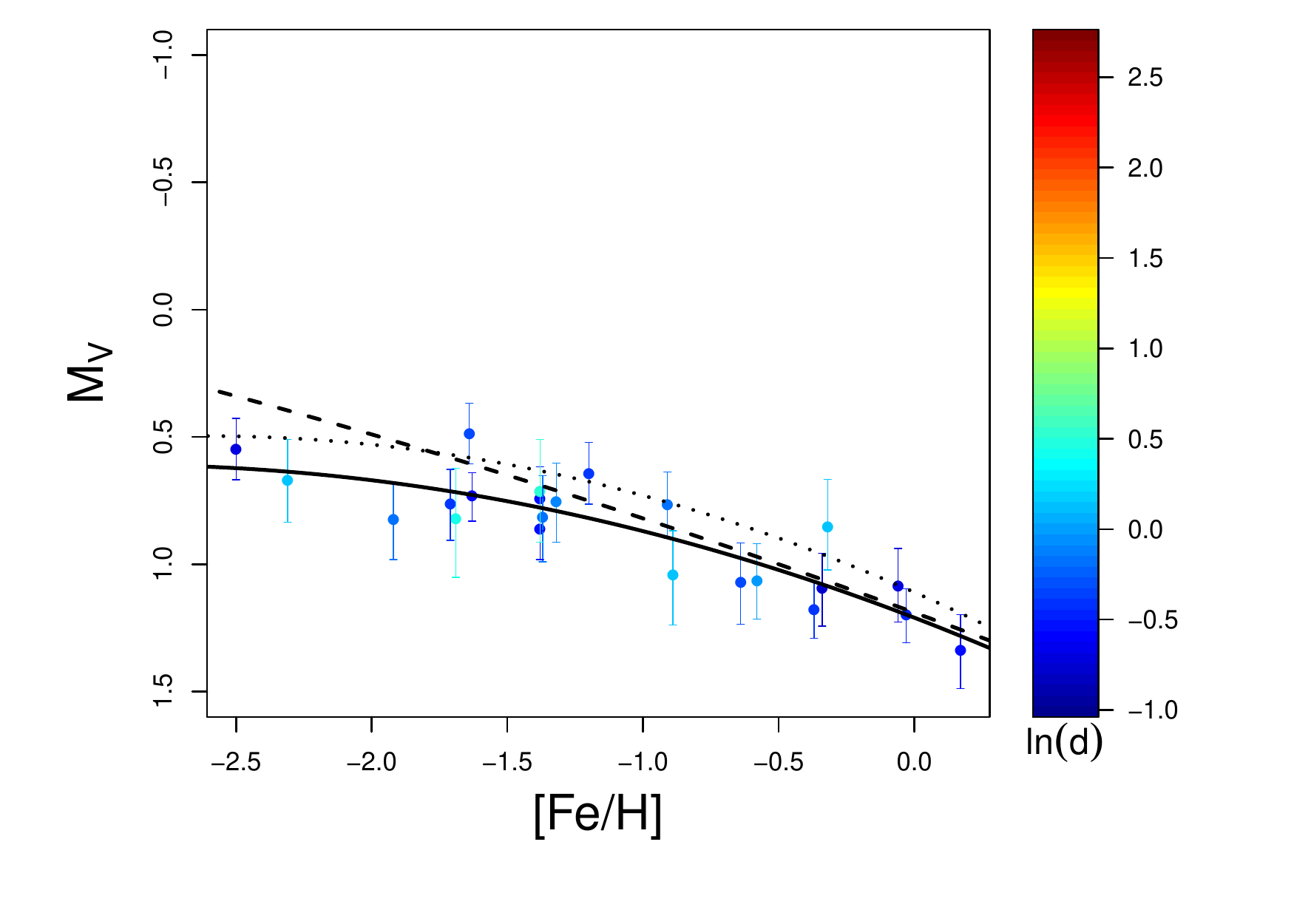}
		\caption{Same as in Fig.~\ref{23-MV-Z-linear-fit} but with lines representing the  quadratic {\mz} relations.}
		\label{23-MV-Z-quadratic-fit}
	\end{center}
\end{figure}

Since a number of theoretical studies suggest a non-linear {\mz} relation, we have also explored quadratic relationships between $M_V$ and
$\rm [Fe/H]$. Table \ref{tab:fittings-DR2} includes a summary of the posterior
distributions of selected model parameters. In the two cases (381 and 23 RRL samples), the
effect of including a second order term is to increase the mean value
of the first order term posterior distribution from 0.34 to 0.39
(sample of 381 stars) and from 0.25 to 0.41 (sample of 23 stars). We mentioned above that there is a controversy relative to the nature (linear or quadratic) of the relationship between metallicity and absolute magnitudes. We have therefore attempted to assess the relative merit of the two models (linear and quadratic) in the light of the available data. In doing so, we are limited by the sampling scheme chosen \citep[Hamiltonian MonteCarlo as implemented in Stan\footnote{Stan is the probabilistic programming language used to code the Bayesian models.};][]{Carpenter2017}. Given this implementation, we cannot use evidences or Bayes' factors and resort to the Bayesian leave-one-out estimate of the expected log-pointwise predictive density  \citep{Vehtari2017}. The comparison of the values obtained for the two models is inconclusive: the mean value of the paired differences is $-4.4\pm
8.8$ favouring the more complex (quadratic) model but with no statistical significance. The quadratic {\mz} relations  for 381 and 23 MW RRLs  are shown in Figs.~\ref{381-MV-Z-quadratic-fit} and \ref{23-MV-Z-quadratic-fit}, respectively.

 We used the relations summarised  in Table~\ref{tab:fittings-DR2} to calculate the mean absolute magnitude of RRLs with metallicity  [Fe/H]=$-1.5$~dex and found values 
of $M_V=0.66\pm0.06$ and $M_V=0.65\pm0.10$ based on the linear {\mz} relations inferred for the full RRL sample and the reduced sample of 23 RRLs with an adopted value of the parallax offset, respectively. These values are in a very good agreement with each other and with the  absolute magnitude found by \citet{Catelan2008} for RR Lyr ([Fe/H]=$-1.48$~dex) $M_V=0.66\pm0.14$~mag.

\begin{table*}
\tiny
		\caption{ RRL {\mz}, $M_{G} - {\rm [Fe/H]}$  and $PMZ$ relations based on {\it Gaia} DR2 parallaxes: (1) relation; (2) number of RRLs used in the fit; (3) mathematical form; (4) intrinsic dispersion; (5) parallax zero-point offset; (6) absolute magnitude in the corresponding passband; (7) distance modulus of the LMC.}
	\begin{center}
		\begin{tabular}{lclccrc}
		\hline
		\hline
		~~~~~~~~Relation 	&No. stars & ~~~~~~~~~~~~~~~~~~~~~~~~~~~~~~~Mathematical form  &  $w$ & $\Delta \varpi_0$ & M$^{*}$~~~~~ & $\mu_{\rm LMC}$ \\
		  & & & (mag) & (mas) & (mag)~~~~ & (mag) \\
			\hline
                           \multicolumn{7}{|c|}{Standard bands}\\
                           \hline
Lin. $M_{V}-[\mathrm{Fe/H}]$   & $381$   & $M_{V}=\left(0.34_{-0.03}^{+0.03}\right)\left[\mathrm{Fe/H}\right]+\left(1.17_{-0.04}^{+0.04}\right)$
			 & $0.14_{-0.01}^{+0.01}$ &	$-0.062_{-0.006}^{+0.006}$ & $0.66\pm0.06$ &  $18.50\pm0.16$ \\
			Quad. $M_{V}-[\mathrm{Fe/H}]$  & $381$   & $M_{V}=\left(0.02_{+0.04}^{+0.04}\right)\left[\mathrm{Fe/H}\right]^{2}+\left(0.39_{-0.09}^{+0.10}\right)\left[\mathrm{Fe/H}\right]+\left(1.19_{-0.06}^{+0.06}\right)$
			 & $0.14_{-0.01}^{+0.01}$ &	$-0.062_{-0.006}^{+0.006}$ & $0.65\pm0.18$ & $18.50\pm0.15$ \\

			$PM_{K_{s}}Z$ 	& $400$ & $M_{K_{s}}=\left(-2.58_{-0.20}^{+0.20}\right)\log\left(P\right)+\left(0.17_{-0.03}^{+0.03}\right)\left[\mathrm{Fe/H}\right]+\left(-0.84_{-0.09}^{+0.09}\right)$ 
			& $0.16_{-0.01}^{+0.01}$ &	$-0.054_{-0.006}^{+0.005}$	& $-0.37\pm0.11$ & $18.55\pm0.11$\\				
			$PM_{W1}Z$  	& $397$ & $M_{W1}=\left(-2.56_{-0.19}^{+0.19}\right)\log\left(P\right)+\left(0.17_{-0.03}^{+0.03}\right)\left[\mathrm{Fe/H}\right]+\left(-0.87_{-0.09}^{+0.09}\right)$   
			 & $0.14_{-0.01}^{+0.01}$ &	$-0.056_{-0.006}^{+0.006}$ & $-0.41\pm0.11$		& $-$ \\ 

                         \hline
			Lin. $M_{V}-\left[\mathrm{Fe/H}\right]$  & $23$   & $M_{V}=\left(0.25_{-0.05}^{+0.05}\right)\left[\mathrm{Fe/H}\right]+\left(1.18_{-0.12}^{+0.12}\right)$ 
			 & $0.13_{-0.03}^{+0.04}$ &	$-0.142_{-0.064}^{+0.058}$   & $0.80\pm0.14$ & $18.34\pm0.16$ \\
			Quad. $M_{V}-\left[\mathrm{Fe/H}\right]$  & $23$ & $M_{V}=\left(0.07_{-0.08}^{+0.07}\right)\left[\mathrm{Fe/H}\right]^{2}+\left(0.41_{-0.18}^{+0.18}\right)\left[\mathrm{Fe/H}\right]+\left(1.21_{-0.13}^{+0.13}\right)$
			 & $0.13_{-0.03}^{+0.04}$ & $-0.127_{-0.066}^{+0.060}$ & $0.75\pm0.35$ &  $18.39\pm0.15$ \\
			$PM_{K_{s}}Z$ 	& $23$ & $M_{K_{s}}=\left(-2.65_{-0.61}^{+0.63}\right)\log\left(P\right)+\left(0.11_{-0.07}^{+0.07}\right)\left[\mathrm{Fe/H}\right]+\left(-0.81_{-0.24}^{+0.26}\right)$
			& $0.12_{-0.03}^{+0.04}$  &	 $-0.135_{-0.063}^{+0.053}$		& $-0.23\pm0.33$ & $18.40\pm0.10$	\\
			$PM_{W1}Z$  	& $23$ &  $M_{W1}=\left(-2.72_{-0.58}^{+0.61}\right)\log\left(P\right)+\left(0.12_{-0.06}^{+0.07}\right)\left[\mathrm{Fe/H}\right]+\left(-0.87_{-0.23}^{+0.25}\right)$
			& $ 0.11_{-0.03}^{+0.04}$ &	$-0.141_{-0.062}^{+0.053} $		& $-0.29\pm0.32$ & $-$ \\

			\hline

            Lin. $M_{V}-[\mathrm{Fe/H}]$   & 23  &  $M_{V}=\left(0.26_{-0.05}^{+0.05}\right)\left[\mathrm{Fe/H}\right]+\left(1.04_{-0.07}^{+0.07}\right)$
			 & $0.13_{-0.03}^{+0.04}$ & $-0.057$		& $0.65\pm0.10$ & $18.49\pm0.15$ \\
			
			Quad. $M_{V}-[\mathrm{Fe/H}]$  & 23  & $M_{V}=\left(0.09_{+0.07}^{+0.07}\right)\left[\mathrm{Fe/H}\right]^{2}+\left(0.47_{-0.16}^{+0.17}\right)\left[\mathrm{Fe/H}\right]+\left(1.11_{-0.09}^{+0.09}\right)$
			& $0.12_{-0.03}^{+0.04}$ & $-0.057$  & $0.61\pm0.31$ & $18.53\pm0.14$ \\
			
                          $PM_{K_{s}}Z$  & 23 &  $M_{K_{s}}=\left(-2.49_{-0.64}^{+0.61}\right)\log\left(P\right)+\left(0.14_{-0.07}^{+0.07}\right)\left[\mathrm{Fe/H}\right]+\left(-0.88_{-0.26}^{+0.25}\right)$ 
			& $0.12_{-0.03}^{+0.04}$  & $-0.057$ & $-0.39\pm0.33$ & 	$18.55\pm0.10$	\\
			
			$PM_{W1}Z$   & 23 & $M_{W1}=\left(-2.54_{-0.58}^{+0.60}\right)\log\left(P\right)+\left(0.15_{-0.06}^{+0.06}\right)\left[\mathrm{Fe/H}\right]+\left(-0.94_{-0.24}^{+0.24}\right)$   & $0.11_{-0.03}^{+0.04}$ 
			& $-0.057$ 	& $-0.45\pm0.31$	&			\\
			
			
			\hline
			                           \multicolumn{7}{|c|}{{\it Gaia} bands}\\
						\hline
			  $M_{G}-[\mathrm{Fe/H}]$ & 160 & $M_{G}=\left(0.32_{-0.04}^{+0.04}\right)\left[\mathrm{Fe/H}\right]+\left(1.11_{-0.06}^{+0.06}\right)$
			  & $0.17_{-0.02}^{+0.02}$ &	$-0.057$	&$0.63\pm0.08$	& $-$					\\ 

			\hline
			
	       \end{tabular}
		\label{tab:fittings-DR2}
	\end{center}
         $^{*}$ Absolute magnitudes of RRLs in different passbands calculated  adopting the metallicity [Fe/H]=$-1.5$~dex and period P=0.5238~days.

\end{table*}


\subsection{Infrared $PMZ$ relations\label{sec:pkz}}

A number of studies on the RRL near-infrared $PM_{K}$ and $PM_{K}Z$  relations exist in the literature both  from the empirical  (see \citealt{Clementini2017} and references therein, for a recent 
historical summary) 
and the theoretical (e.g. 
\citealt{Marconi2015} and references therein) points of view. 
While empirical studies suggest a mild or even negligible dependence of the $K$-band luminosity on metallicity, the theoretical studies find for the metallicity term of the $PM_{K}Z$ relation
slope values  up to $0.231\pm0.012$ \citep{Bono2003}.  The literature values for the dependence of the $M_K$ magnitude on period also vary ranging from $-2.101$ \citep{Bono2003} to $-2.73\pm0.25$ (see table~3 in \citealt{Muraveva2015}, for a compilation).

The RRL mid-infrared relations at the $W1$ (3.4 $\mu$m) passband of WISE, $PM_{W1}$ and $PM_{W1}Z$,  have also been studied by many different authors both on empirical  (e.g. 
 \citealt{Sesar2017}, \citealt{Clementini2017}, and references therein) and theoretical (e.g \citealt{Neeley2017}) grounds,  with theoretical studies suggesting a non-negligible dependence on metallicity of $0.180\pm0.003$ mag/dex, (\citealt{Neeley2017}).  For comparison, \citet{Dambis2014}  derived  a dependence on metallicity of $0.096\pm0.021$ mag/dex of the $PM_{W1}Z$ relation,  
from their studies of RRLs in globular clusters, while \citet{Sesar2017} derived a  metallicity slope of $0.15\substack{+0.09\\-0.08}$  mag/dex using TGAS parallaxes for a sample of about a hundred  MW RRLs.  The literature values of the period slope vary from $-2.15\pm0.23$ (3.6 $\mm$ passband; \citealt{Muraveva2018}) to  $-2.47\pm0.74$ \citep{Sesar2017}.

\begin{figure}
	\begin{center}
		\hspace*{-0.8cm}
		\includegraphics[trim=0 40 60 10, width=\linewidth]{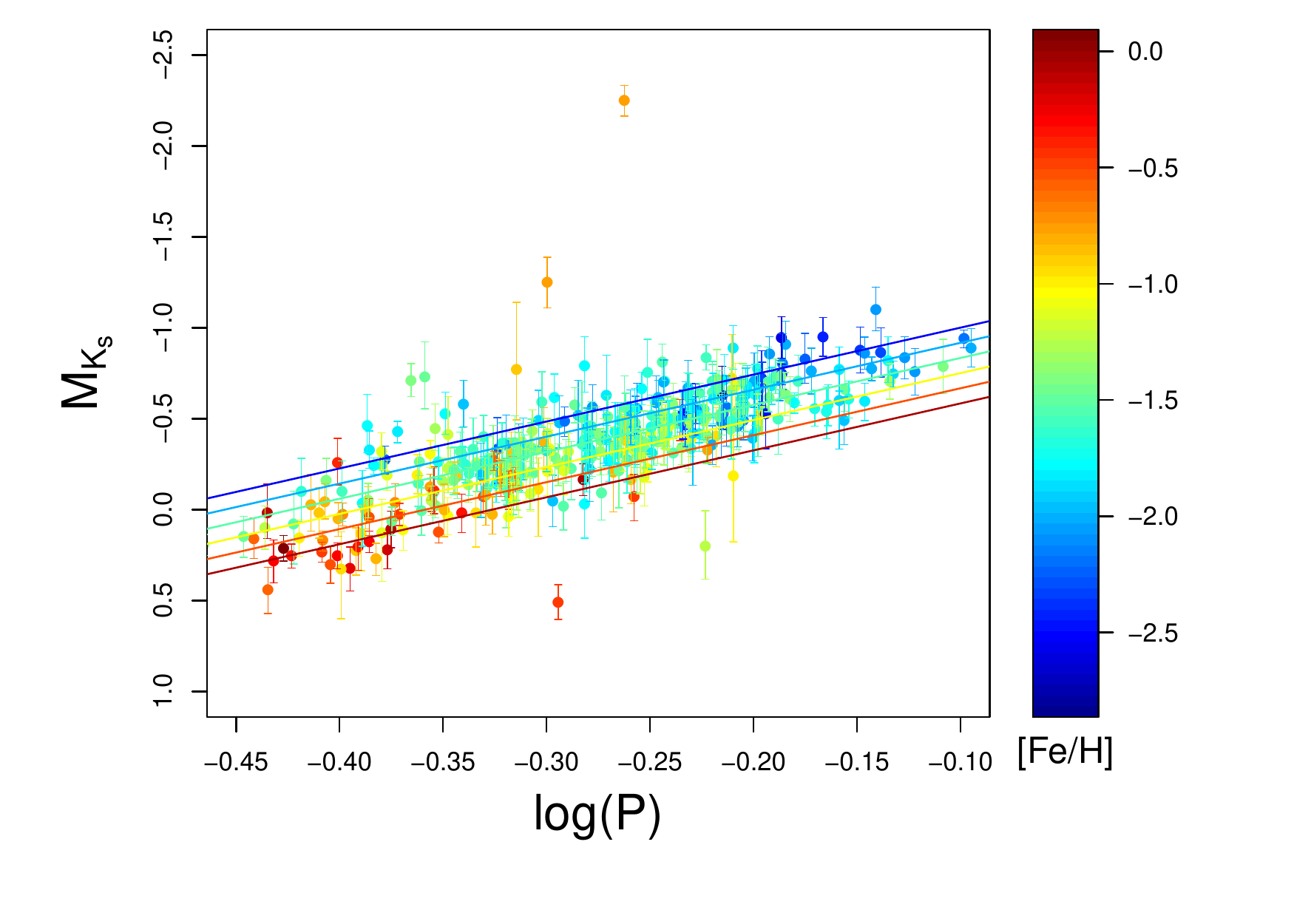}
		\caption{$PM_{K_s}$  distribution of 400 RRLs in our sample,
			for which absolute $M_{K_s}$ magnitudes were inferred from the model
			described in Section \ref{sec:met}. The lines represent projections of the 
			fit shown in the first portion of Table \ref{tab:fittings-DR2} onto the Magnitude-Period
			plane. The colour scale encodes metallicity values measured  on the \citet{ZW} metallicity scale.}
		\label{396-MKZ-fit}
	\end{center}
\end{figure}

\begin{figure}
	\begin{center}
		\hspace*{-0.8cm}
		\includegraphics[trim=0 40 60 20, width=\linewidth]{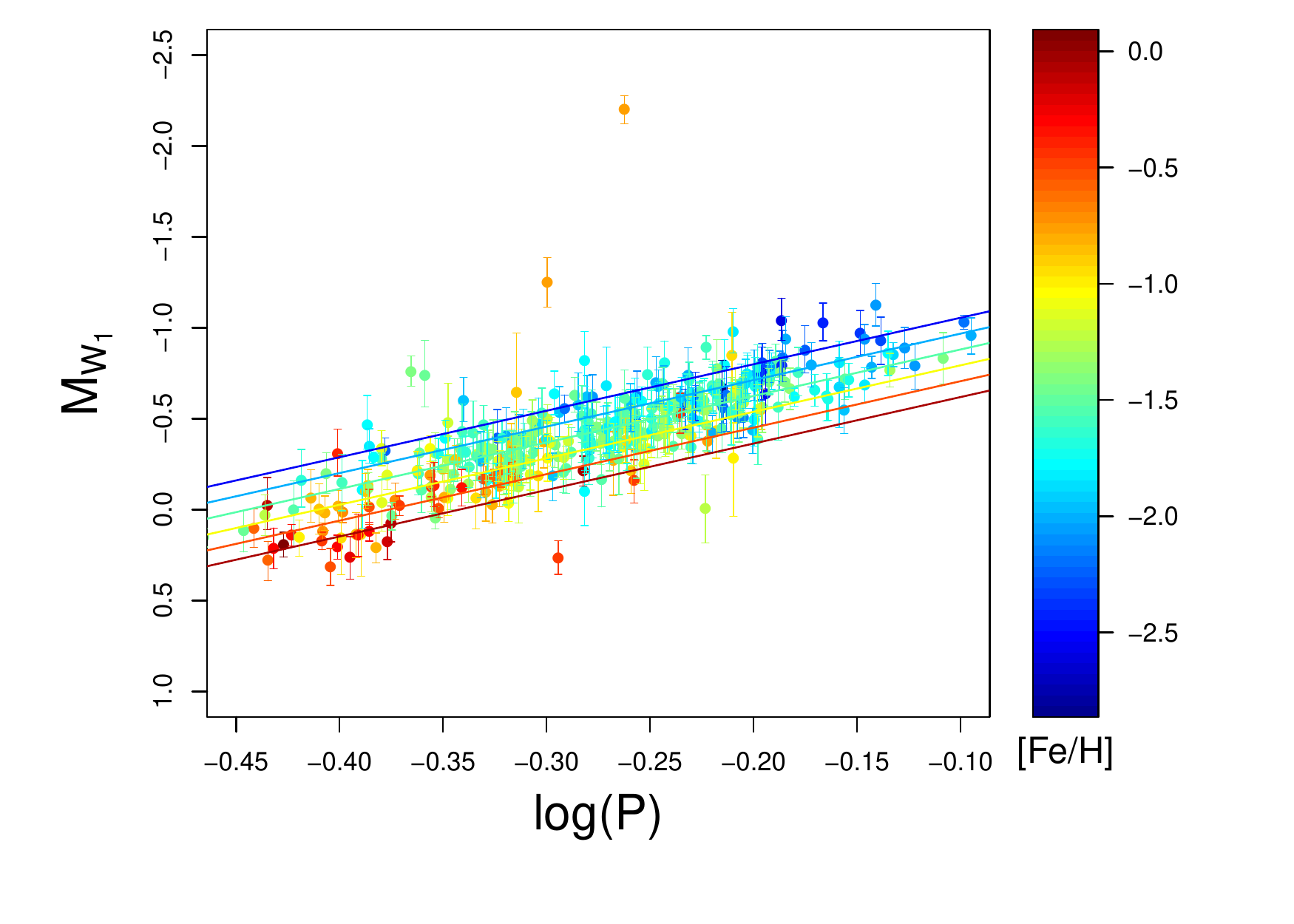}
		\caption{$PM_{W1}$  distribution of 397 RRLs from our sample,
			for which absolute $M_{W1}$ magnitudes were inferred from the model
			described in Section \ref{sec:met}. The lines represent projections of the 
			fit shown in the first portion of Table \ref{tab:fittings-DR2} onto the Magnitude-Period
			plane. The colour scale encodes  metallicity values measured on the \citet{ZW} metallicity scale.  }
		\label{397-MW1Z-fit}
	\end{center}
\end{figure}

\begin{figure}
	\begin{center}
		\hspace*{-0.8cm}
		\includegraphics[trim=0 40 60 10, width=\linewidth]{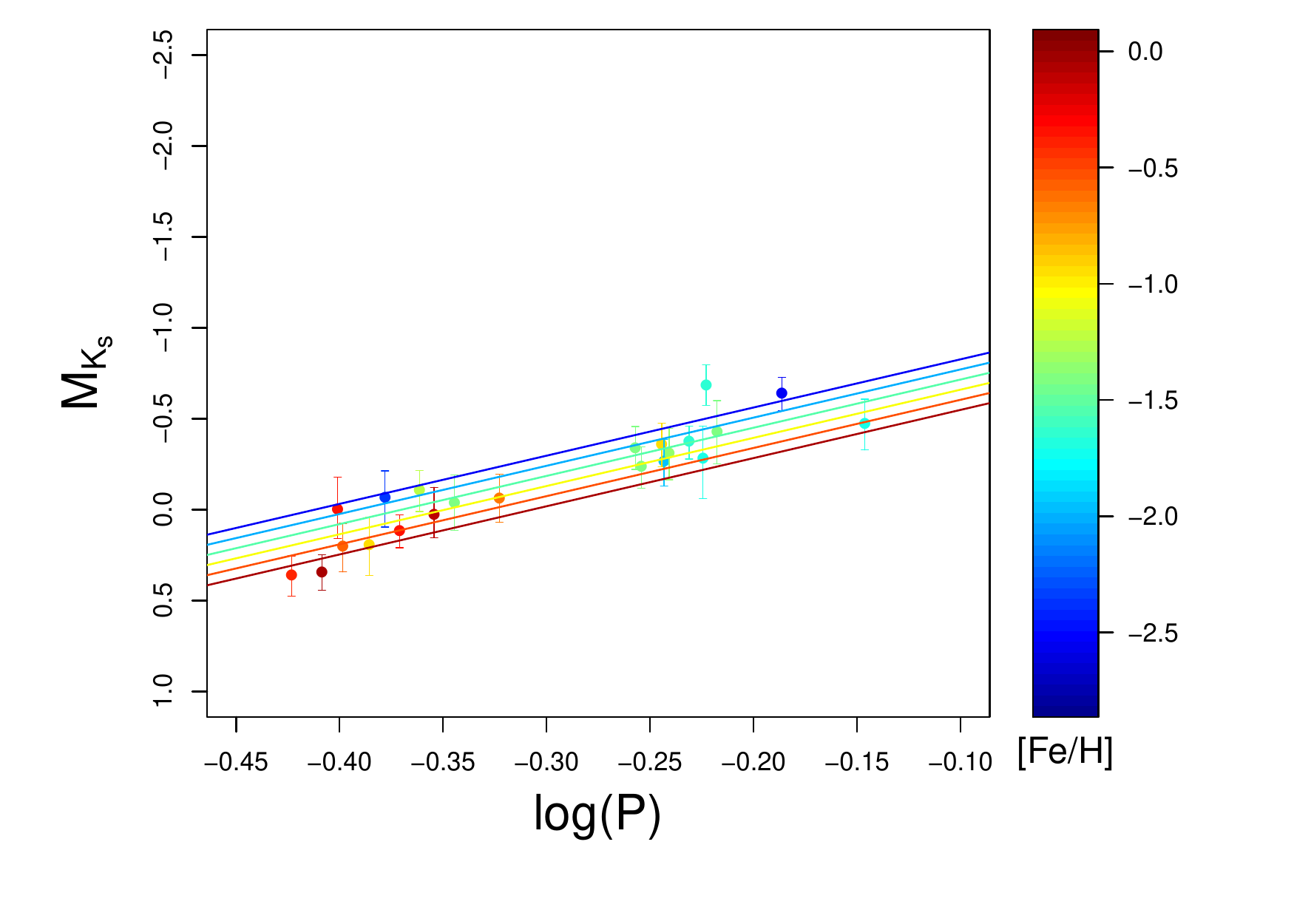}
		\caption{$PM_{K_s}$  distribution of 23 RRLs from our sample,
			for which absolute $M_{K_s}$ magnitudes were inferred from the model
			described in Section \ref{sec:met}. The lines represent projections of the 
			fit shown in the second portion of Table \ref{tab:fittings-DR2} onto the Magnitude-Period
			plane. The colour scale encodes   metallicity values measured  on the high-resolution spectroscopy metallicity scale.}
		\label{23-MKZ-fit}
	\end{center}
\end{figure}

\begin{figure}
	\begin{center}
		\hspace*{-0.8cm}
		\includegraphics[trim=0 40 60 20, width=\linewidth]{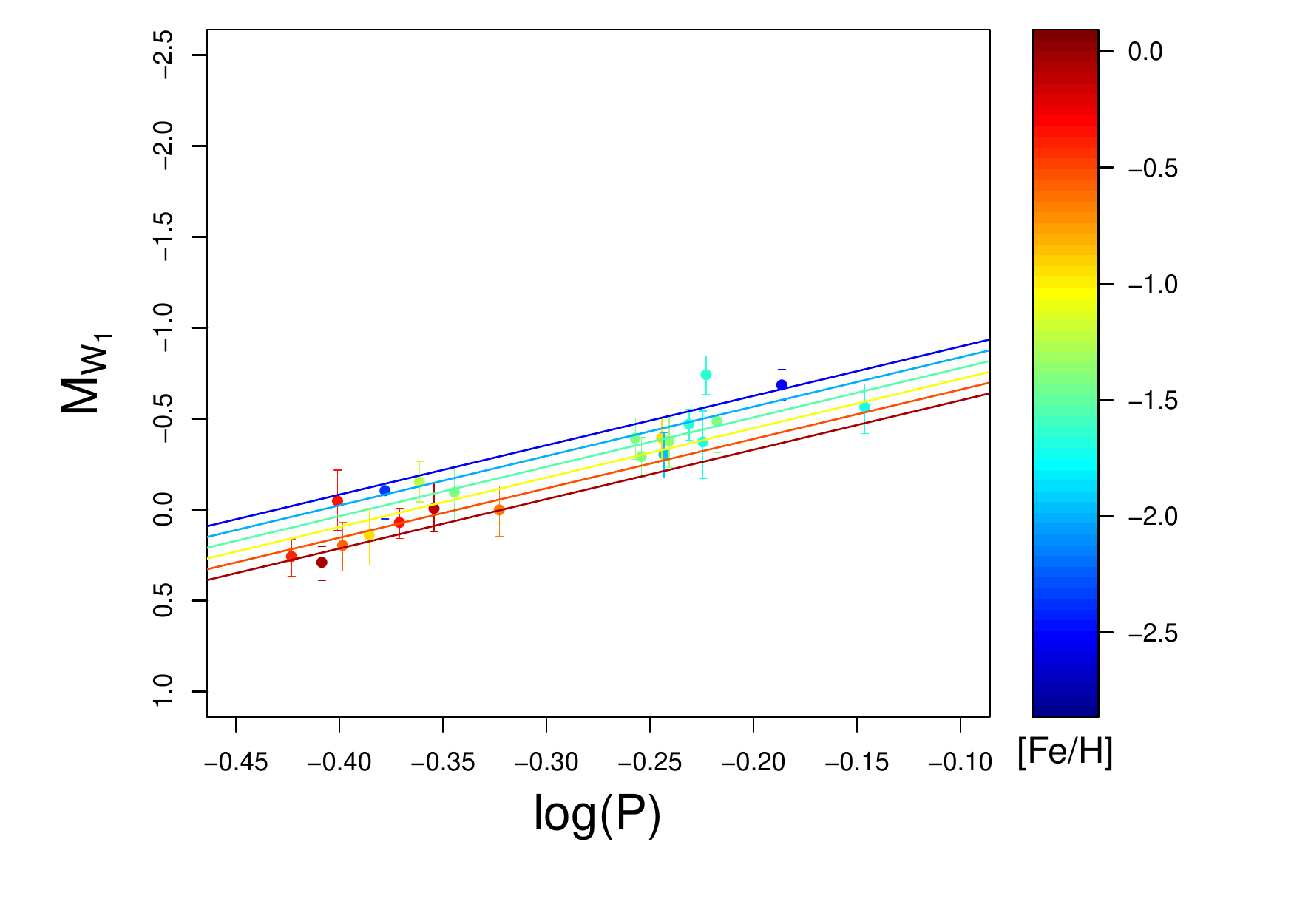}
		\caption{$PM_{W1}$  distribution of 23 RRLs from our sample,
			for which absolute $M_{W1}$ magnitudes were inferred from the model
			described in Section \ref{sec:met}. The lines represent projections of the 
			fit shown in the second portion of Table \ref{tab:fittings-DR2} onto the Magnitude-Period
			plane. The colour scale encodes  metallicity values measured on the high-resolution spectroscopy metallicity scale. }
		\label{23-MW1Z-fit}
	\end{center}
\end{figure}

We derived infrared  $PM_{K_{\rm s}}Z$ and $PM_{W1}Z$ relations for the RRLs in our sample using the Bayesian approach described in Section~\ref{sec:met} and inferring the parallax zero-point offset from the model.  The near-infrared $PM_{K_{\rm s}}Z$ relation  is  based on  a sample of 400 RRLs for which all needed information along with apparent $K_{\rm s}$ magnitudes and related  uncertainties are available in 
\citet{Dambis2013} (see Section~\ref{sec:data}).  
To derive the $PM_{W1}Z$ relation we used a sample of 397 RRLs for which $W1$ magnitudes and related uncertainties are also available in \citet{Dambis2013}.
The coefficients of the resulting relations are summarised in the first section of Table~\ref{tab:fittings-DR2} (row 3 and 4 respectively) and graphically  shown in Figs.~\ref{396-MKZ-fit} and \ref{397-MW1Z-fit}, where the colours encode the RRL metallicities on the \citet{ZW} metallicity scale. The slope in period we derive for the  {\pkz} relation is in perfect agreement with the literature values, while the metallicity slope is higher than found in previous empirical studies  but in excellent agreement with the theoretical findings (e.g \citealt{Bono2003}, \citealt{Catelan2004}). The slope in period  of the {\pwz} relation  is slightly steeper than the literature values. We also find a non-negligible metallicity dependence, that is consistent  with results from \citet{Neeley2017} and \citet{Sesar2017}.  The mean value of the parallax offset derived from  fitting  the linear {\mz} (Section~\ref{sec:mz}) and the {\pkz}, {\pwz} relations of the full sample of RRLs is $-0.057$~mas, which is in very good agreement with the offset value  found  for RRLs by \citet{Arenou2018}. In particular, the offset inferred from the model for the {\pwz} relation matches exactly the \citet{Arenou2018}'s value.  

We have  performed the fitting on our reduced sample of 23 RRLs both inferring the parallax offset as a parameter of the model and assuming a  constant offset value of  $-0.057$~mas. The resulting relations are presented in the second and third portions  of Table~\ref{tab:fittings-DR2}.  Figs.~\ref{23-MKZ-fit} and \ref{23-MW1Z-fit} graphically show the {\pkz} and {\pwz} relations obtained from this reduced sample of 23 MW RRLs, when inferring the parallax offset as a parameter of the model. As with the RRL {\mz} relation  the metallicity slope is significantly shallower for the reduced sample of 23 RRLs and in agreement within the uncertainties with values presented in the literature (e.g. \citealt{Catelan2004}, \citealt{Neeley2017}, \citealt{Sesar2017}). 
 As in Section~\ref{sec:zp} we calculated the mean $M_{K_{\rm s}}$ and $M_{W1}$ absolute magnitudes of an RRL with metallicity  [Fe/H]=$-1.5$~dex and  period  P=0.5238~days, which is the 
mean period of the RRLs in our sample. The resulting values are presented in column 6 of Table~\ref{tab:fittings-DR2}.

\begin{figure*}
	\begin{center}
		\includegraphics[scale=0.4]{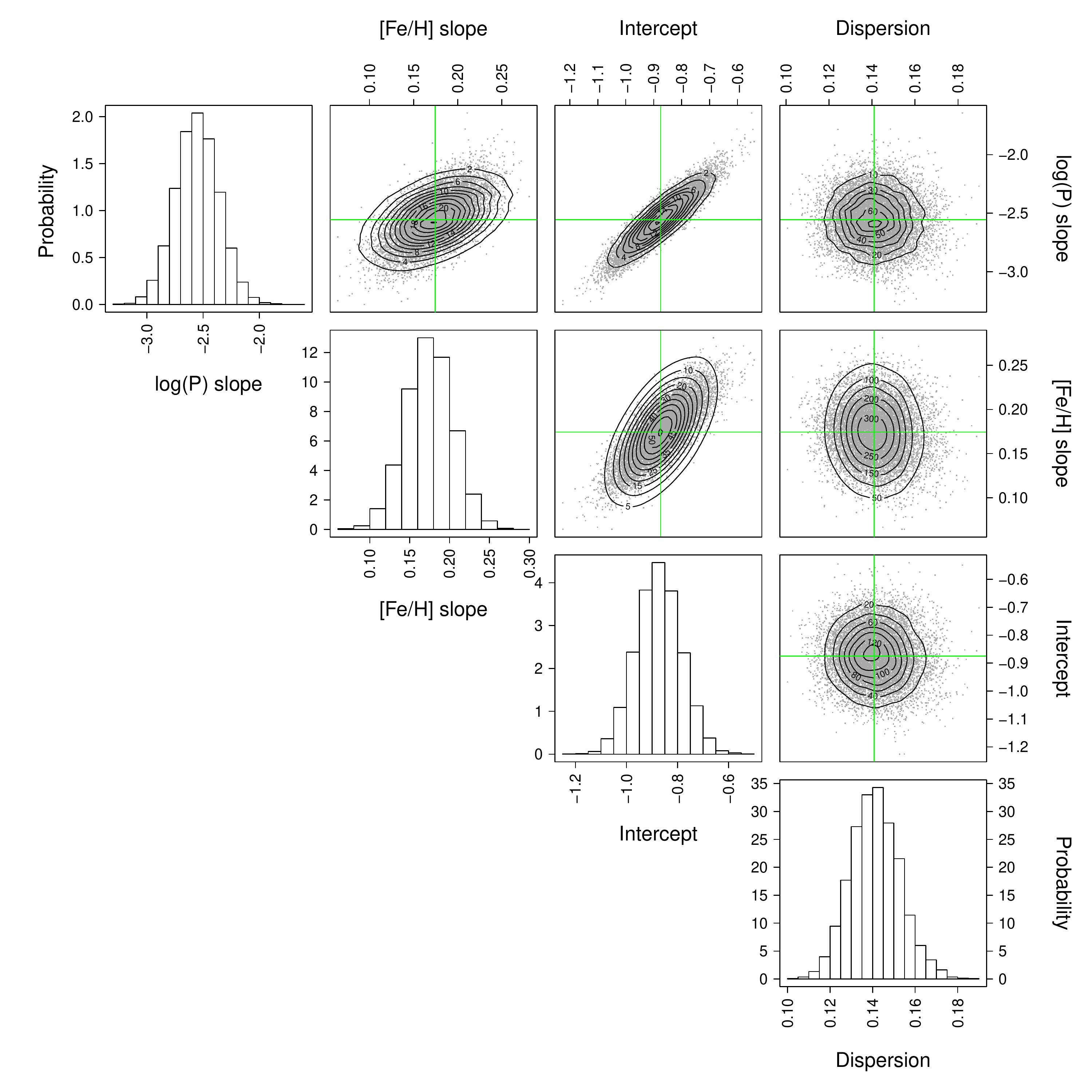}
		\caption{Marginal posterior distributions of the $W1$-band dataset in different one- and
                  two-dimensional projections: from left to right, the
                  $\log(P)$ and $\left[\mathrm{Fe/H}\right]$ coefficients, the intercept
                  and the intrinsic dispersion. Each green line points towards the median 
                  of the corresponding one-dimensional posterior marginal distribution.}
		\label{fig:posterior-samples}
	\end{center}
\end{figure*}

 Fig.~\ref{fig:posterior-samples} shows the marginal posterior distributions in different one- and two-dimensional projection planes for the $W1$-band dataset. These distributions are representative/qualitatively similar to those obtained for the other models discussed in this study.

\subsection{$M_{ G} - {\rm [Fe/H]}$ relation}
 
\begin{figure}
	\begin{center}
		\hspace*{-0.8cm}
		\includegraphics[trim=0 40 60 10, width=\linewidth]{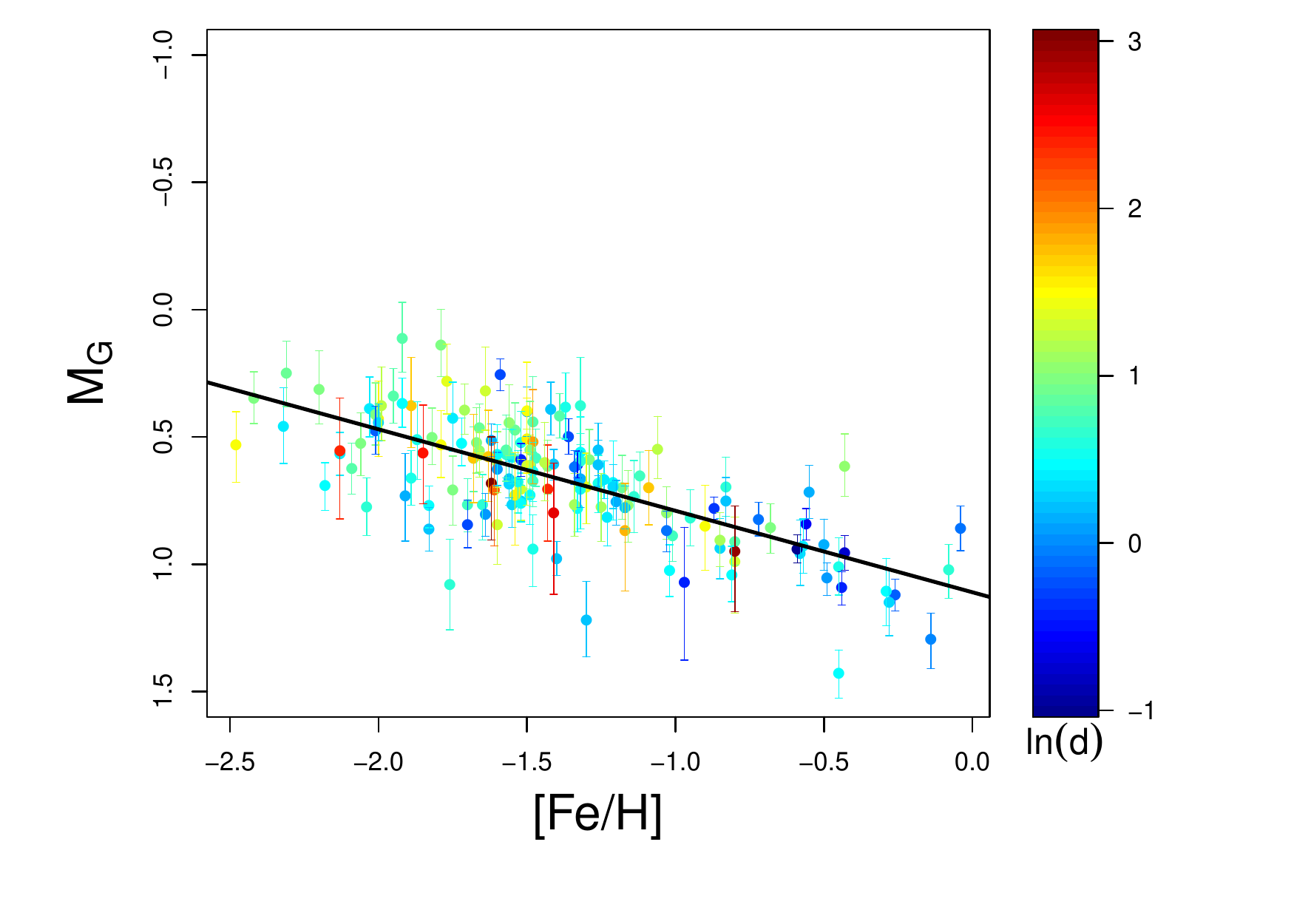}
		\caption{$M_G - {\rm [Fe/H]}$ relation defined by 160 RRLs  in our sample,
			whose absolute $M_G$ magnitudes were inferred from the model
			described in Section \ref{sec:met}. The solid line represents the linear fit. Its slope and zero-point 
			are summarised in the last portion of Table \ref{tab:fittings-DR2}. The colour scale 
			encodes the (natural) logarithm of the inferred (true) distance measured in units of kpc.}
		\label{160-G-Z}
	\end{center}
\end{figure}

The {\gaia} DR2 catalogue contains magnitudes in the {\gaia} $G$-band ($330 - 1050$~nm) for $\sim1.7$ billion sources and $G_{BP}$, $G_{RP}$ photometry derived from the integration of the blue and red photometer low-resolution spectra ($G_{BP}$: $330 - 680$~nm; $G_{RP}$: $630 - 1050$~nm) for $\sim1.4$ billion sources \citep{Evans2018}. 
For sources confirmed to be RRLs {\gaia} DR2 also published  intensity-averaged mean $G$, $G_{BP}$, $G_{RP}$ magnitudes computed by modelling the multi-band light curves over the whole pulsation cycle  and extinction values in the $G$-band inferred from the RRL pulsation characteristics \citep{Clementini2018}. Specifically, intensity-averaged mean $G$ magnitudes are available for 306 of the RRLs in our sample and the $G$-band extinction values are available for 160 of them.
We used the sample of 160 RRLs along with their metallicities from \citet{Dambis2013} and our Bayesian model with an adopted parallax offset of $-$0.057 mas to fit the RRL  $M_{ G} - {\rm [Fe/H]}$ relation. The relation is shown in 
the last portion of Table~\ref{tab:fittings-DR2} and in Fig.~\ref{160-G-Z}. 
The corresponding RRL $G$-band absolute magnitude at [Fe/H]=$-1.5$~dex is  $M_G=0.63\pm0.08$~mag.  
This value is consistent with the $V$-band absolute magnitudes derived in Section~\ref{sec:zp} and can be used to infer an approximate estimation of distance to RRLs whose  apparent mean magnitude and extinction in the $G$-band are available in the {\gaia} DR2 catalogue.

%

\section{Distance to the LMC} \label{sec:lmc}

\begin{figure}
	\begin{center}
                \includegraphics[trim=70 170 250 215, width=\linewidth]{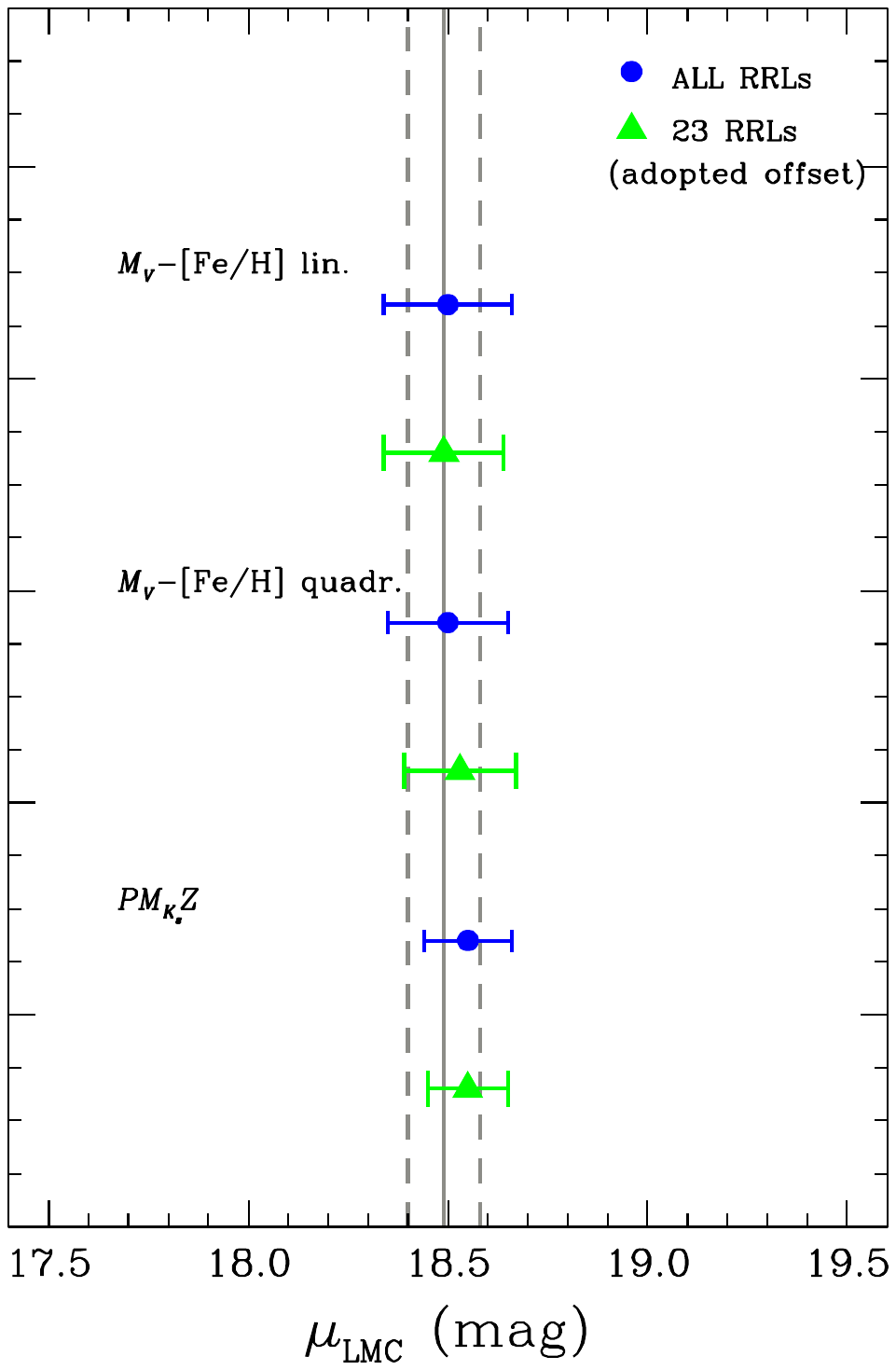}
		\caption{Distance moduli of the LMC obtained from the  {\mz} and  {\pkz} relations derived  in this paper 
		 using the full sample of MW RRLs (blue filled circles; values in column 7 of the  first portion  of Table~\ref{tab:fittings-DR2})  and the sample of 23 MW RRLs  with an assumed constant value of the parallax offset of $-0.057$~mas (green triangles; values in column 7 of the third  portion of Table~\ref{tab:fittings-DR2}). See text for the details.}
		\label{fig:lmc}
	\end{center}
\end{figure}

As traditionally done in this type of studies, in order to test the  {\it Gaia} DR2 parallax-calibrated relations of RRLs derived in  Section~\ref{sec:relations}
we apply them to infer the distance to the LMC,  a cornerstone of the cosmological distance ladder, whose distance has been measured in countless studies with different distance indicators and independent techniques. 
Following \citet{Clementini2017},  we considered 
70 RRLs located  close to the LMC bar, for which spectroscopically measured metallicities \citep{Gratton2004}, extinction, periods and photometry in the $V$ \citep{Clementini2003} and $K_{\rm s}$ \citep{Muraveva2015} bands, are available. 
No $W1$-band photometry is available for these 70 LMC RRLs, while intensity-averaged $G$ mean magnitudes are available for 44 of them. However,  the $G$-band extinction values  
are  available only for two of the stars in this sample.  Hence, we only  considered the $V$,  $K_{\rm s}$ magnitudes  and 
applied the linear and quadratic {\mz} relations  and the {\pkz} relations in Table~\ref{tab:fittings-DR2}, to infer  the distance to each RRL individually and then computed the weighted mean of the distribution.
The metallicity scale adopted by \citet{Gratton2004}, is 0.06~dex more metal-rich than the \citet{ZW} metallicity scale. We subtracted 0.06~dex from the metallicities of the LMC RRLs to convert them to the \citet{ZW} metallicity scale when dealing with the relations based on the whole sample of RRLs. No correction was applied instead 
when using  the relations based on the 23 MW RRLs with metal abundances obtained from high-resolution spectroscopy.  
The LMC distance moduli obtained with this procedure  are summarised in the last column of Table~\ref{tab:fittings-DR2} and plotted in Fig.~\ref{fig:lmc},  where they are shown  to agree within 
1 $\sigma$ uncertainty (grey dashed lines)  with the very precise LMC modulus: $\mu= 18.493 \pm 0.008 ({\rm stat}) \pm 0.047 ({\rm syst}$) mag inferred  by \citet{Pietr2013} from the analysis of eight eclipsing binaries in the LMC bar (grey solid line).
 
 We do not plot in Fig.~\ref{fig:lmc} distance moduli obtained from the relations defined by the sample of 23 RRLs and the parallax offset  inferred from the model, because the offset  is significantly overestimated ($-$0.142 mas) in this case and the corresponding moduli underestimated.   
On the other hand, the relations based on the sample of 23 MW RRLs and an assumed constant value of the parallax offset of $-0.057$~mas produce LMC distance moduli (green triangles)  in very good agreement with the canonical value by \citet{Pietr2013}. To conclude, the LMC distance moduli obtained in this study using the {\it Gaia} DR2 parallaxes are in good agreement with the canonical value,
once the {\it Gaia} DR2 parallax offset is properly accounted for.

\section{Summary and conclusions} \label{sec:summ}
{\it Gaia}  Data Release 2 provides accurate parallaxes for an unprecedented, large number of MW RRLs.  In this study we analysed a sample of 401 MW field RRLs, for which $V$, $K_\mathrm{s}$, and $W1$ photometry, metal abundances, extinction values and pulsation periods are available in  the literature and accurate parallaxes  have become available with the {\it Gaia} DR2.
We compared the {\gaia} DR2 parallaxes with the parallax estimates for these RRLs available in the Hipparcos, {\it HST} and TGAS catalogues.
We find a general good agreement of the  {\it Gaia} DR2 parallaxes with the TGAS and  the {\it HST} measurements, while agreement with the Hipparcos catalogue is less pronounced. The accuracy of the DR2  parallax measurements for RRLs showcases an  impressive improvement achieved by {\it Gaia} both with respect to its predecessor Hipparcos and the TGAS measurements released in DR1, and rivals to other space-born estimates by cutting down about a factor of 5 the parallax uncertainty for RRLs measured with the {\it HST}.

With {\it Gaia}  DR2 it is for the first time possible to determine the coefficients (slopes and zero-points)  of the fundamental relations ({\mz}, {\pkz}, {\pwz},  as well as the {\it Gaia}  $M_G - {\rm [Fe/H]}$ relation),
that RRLs  conform to on the basis of statistically significant samples of stars with accurate parallax measurements availble, that we do in this paper by applying a fully Bayesian approach that properly handles parallax measurements and biases affecting  our sample of 401 MW RRLs. We find the dependence of the luminosity on metallicity to be higher than usually adopted in the literature. We show that this high metallicity dependence is not caused by our inference method, but likely arises from the actual distribution of the data and it is strictly connected with a possible offset affecting  the {\it Gaia} DR2 parallaxes.  
This effect is much reduced for a sample of 23 MW RRLs with the metallicity  estimated from high-resolution spectroscopy, which are closer to us and span a narrower range of the distances. However, we caution the reader that selection effects can potentially be stronger for nearby RRLs. Using our Bayesian approach we recover an offset of about $-0.057$~mas affecting  the {\it Gaia} DR2 parallaxes of our full sample of about 400 RRLs, confirming previous findings by \citet{Arenou2018}. 

Our study demonstrates the effectiveness of the {\it Gaia} parallaxes to establish the cosmic distance ladder by recovering the canonical value of 18.49 mag for the distance modulus of the LMC, once the DR2 parallax offset is properly corrected for. We hence confirm that {\it Gaia} is on the right path and look forward to  DR3, which is currently foreseen for end of 2020.

\section*{Acknowledgements}

This work makes use of data from the ESA mission {\it Gaia} (\url{https://www.cosmos.esa.int/gaia}), processed by
the {\it Gaia} Data Processing and Analysis Consortium (DPAC, \url{https://www.cosmos.esa.int/web/gaia/dpac/consortium}). 
Funding for the DPAC has been provided by national institutions, in particular
the institutions participating in the {\it Gaia} Multilateral Agreement.
Support to this study has been provided by PRIN-INAF2014, "EXCALIBUR'S" (P.I. G. Clementini), 
from the Agenzia Spaziale Italiana (ASI) through grants ASI I/058/10/0 and ASI 2014-
025-R.1.2015 and by Premiale 2015, ``MITiC" (P.I. B. Garilli).
We thank Prof.  J. Lub for useful updates on some of the entries in the catalogue of RRLs used in this study. 
The statistical analysis carried out in this work has made extensive use of the {\sl R} statistical software and, in particular, the {\sl Rstan} package.





\bsp	
\label{lastpage}
\end{document}